\documentstyle[aaspp4,rotating,psfig]{article}

\begin{document}
\title {On the Nature of Pulsar Radio  Emission}
\author{ Maxim Lyutikov \footnote{Currently at the Canadian Institute
for Theoretical Astrophysics, 60 St. George, Toronto, Ont,  M5S 3H8, Canada},
 Roger D. Blandford }
\affil{Theoretical Astrophysics, California Institute of Technology,
Pasadena, California 91125, USA}
\author{George Machabeli}
\affil{
Abastumani Astrophysics Observatory, A. Kazbegi Av. 2a, Tbilisi, 380060
Republic of Georgia} 
\date { draft}

\begin{abstract}
A theory of pulsar radio emission generation,
in which the observed waves  are produced directly
 by   maser-type
plasma instabilities operating at the anomalous cyclotron-Cherenkov resonance
$\, \omega-\, k_{\parallel} v_{\parallel} +\, \omega_B/\,  \gamma_{res}=0$
and the  Cherenkov-drift resonance
$\, \omega-\, k_{\parallel} v_{\parallel} - k_{\perp} u_d =0$,
 is capable of explaining the main observational
characteristics of pulsar radio emission. The
instabilities are due to the interaction of the fast particles from
the primary beam and  the tail of the distribution
 with the  normal modes of a strongly magnetized one-dimensional
electron-positron plasma.
The waves emitted at these resonances are vacuum-like,  electromagnetic
waves that may leave the  magnetosphere directly.
In this model, 
the cyclotron-Cherenkov instability  is responsible for
 core emission pattern and
the Cherenkov-drift instability produces conal  emission.
The conditions for the development of the cyclotron-Cherenkov
instability are satisfied for  both typical and millisecond
pulsars provided that the streaming energy of the bulk
plasma is not very high $\, \gamma_p \approx 10$.  In a typical pulsar
the cyclotron-Cherenkov and Cherenkov-drift
 resonances occur in the outer parts of
magnetosphere at $ r_{res}\, \approx 10^9 {\rm cm}$.
This theory can account for various aspects of pulsar phenomenology
including
the morphology of the pulses, their polarization  properties and their 
spectral behavior.
We propose several observational tests for the theory. The most prominent
prediction are the high altitudes of the emission region
and the linear polarization  of conal emission in the 
plane {\it orthogonal} to the local osculating plane of the magnetic field.

\end{abstract}


\section{Introduction}

More than twenty five years have passed since the discovery of pulsars
and  there is still  no consensus  on the basic emission mechanism.
At the present time, there  are about a dozen competing theories which differ
both in the physical effects responsible for the 
 radiation and in the locations where
they operate (\cite{Melrose-DB}).
 Probably the only point of agreement between all these theories
is the association of pulsars with magnetized, rotating neutron stars.
By contrast, there is so much  observational
data available that  none of the existing theories can  explain
all the {\it main}
observational  facts.

There are  several reasons that have precluded understanding of  pulsar
radio emission. First, there is the unusual physical conditions
found in pulsar magnetospheres (relativistic electron-positron plasma,
superstrong magnetic fields, ultrarelativistic beam).
Secondly,  only a small fraction of  the energy lost by a 
neutron star is re-emitted in the  radio (even in the high energy
range, where a considerable portion of energy is emitted, 
there is still no consensus on the origin of this emission, 
e.g.,  \cite{DaughertyHarding1996}, \cite{Romani1996}).
The third reason is that  after thirty years of research, we still 
do not know the general structure of a pulsar magnetosphere
(\cite{Mestel1995}). We understand  only  particular features,
 like the  existence of open and closed field lines and 
where  electric field  can be parallel to the 
magnetic field.

To date,
the most widely discussed  theory attributes the emission to 
 coherent curvature  emission
by bunches of particles. Though this theory can
explain a broad
range of observed pulsar  properties by the  careful arrangement of the
magnetic field
geometry and of the form and size of bunches,  thirty years of 
theoretical efforts have failed to explain the origin of these bunches
(\cite{Melrose-DB}).  This theory can also be ruled out on the observational
grounds (\cite{Lesch}). In addition to the work of \cite{Lesch} we note, that
this theory also fails to explain 
 the observed correlations of the conal peaks  \cite{Smirnova1991}
and a large  size of the emitting region  \cite{Gwinn}
(see Section \ref{Observational properties}).

We propose that the pulsar radiation is generated by {\it plasma
instabilities} developing in the outflowing plasma on the open
field lines
 of the  pulsar
 magnetosphere. Plasma
can be considered as an active medium that can amplify its
normal modes. In the case of the two instabilities discussed  below,
the wave amplification is due to the  {\it resonant} wave-particle
interaction, i.e., in the rest frame of the particle the frequency of the
resonant wave is zero or a multiple of the gyrational frequency. 
The plasma instabilities that we argue operate in the pulsar
magnetosphere may be described by the (somewhat contradictory) term
"incoherent broadband maser". Each single emission by a charged
particle is due to the
stimulated, as opposed to spontaneous,
 emission process (hereby the term maser).
Unlike the conventional lasers in which basically one single frequency 
gets amplified, in this case charged particles can resonate 
with many mutually incoherent waves  with different frequencies.

In this paper we discuss a  theory of pulsar radio emission
developed by \cite{LominadzeMachabeliMkhailovsky},
 \cite{MachabeliUsov1989},
 \cite{Kazbegi}, \cite{Machabeli2},
 \cite{LyutikovDisp},
\cite{LyutikovMachabeliBlandford1}.
We hypothesize, that pulsar radiation is generated by the 
instabilities developing in the outflowing plasma on the open
field lines in the outer
regions of the  pulsar  
 magnetosphere. 
Radiation is generated by two kinds of 
{\it electromagnetic}  plasma instabilities --
cyclotron-Cherenkov and Cherenkov-drift instabilities. The
cyclotron-Cherenkov instability is responsible for the
generation of the core-type emission and the 
Cherenkov-drift instability is responsible for the
generation of the cone-type emission (\cite{rankin1}).
The waves generated by these instabilities
are  vacuum-like electromagnetic waves so that  
they may leave magnetosphere the
directly.

In contrast to most modern theories of pulsar radio emission, 
cyclotron-Cherenkov and Cherenkov-drift
instabilities occur in the outer parts of magnetosphere.
The location of the
emission region is determined by the corresponding
resonant condition for the cyclotron-Cherenkov and
Cherenkov-drift instabilities. Instabilities develop in a limited
region on the open field lines. The size of the emission
region is determined by the curvature of the magnetic
field lines, which limits the length of the resonant wave-particle
interaction. The location of the cyclotron-Cherenkov
instability is restricted to those field lines with large radius
of curvature,
while the Cherenkov-drift instability occurs on  field lines
with   curvature
bounded  both from above and from below.
Thus, both instabilities produce narrow pulses, though they
operate at radii where the opening angle of the open field  lines
is large.

\begin{figure}
\psfig{file=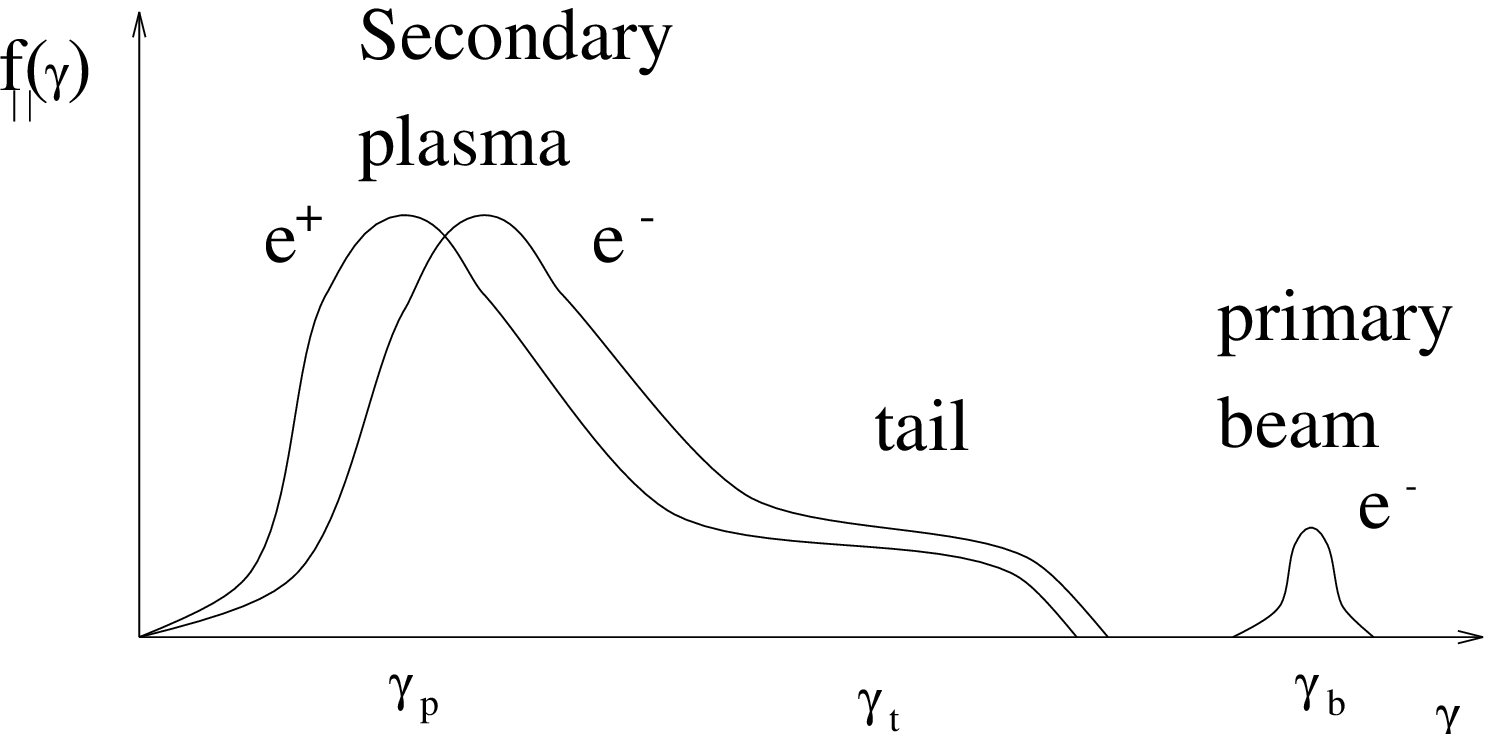,width=15.0cm}
\caption{
 Distribution function for a one-dimensional electron-positron
plasma of pulsar magnetosphere.
\label{Distributionfunction}}
\end{figure}

To a large extent a possible mechanism for the generation of pulsar radio
emission is predicated on the choice of parameters of the plasma flow that
is generated by  a rotating neutron star.
At this point we know
only the general features of the 
distribution function of the particles in a pulsar magnetosphere
(\cite{Tademaru},\cite{Arons1981}, \cite{DaughertyHarding1983}).
It is believed to comprise (see Fig. \ref{Distributionfunction}) 
(i) a highly relativistic primary beam
with the Lorentz factor $\,  \gamma_b \approx 10^7$ and 
density equal to the Goldreich-Julian density $n_{GJ}={\bf \Omega \cdot B}/(2\,
\pi\, e\, c)$, (ii)   a secondary 
electron positron plasma with a bulk streaming Lorentz factor
$\,  \gamma_p \approx 10 -1000$, a similar  scatter in energy 
$T_p \approx \,  \gamma_p$ and a density much larger
than  the beam density $n_p \approx \, \lambda\, n_{GJ} =10^3-10^6 n_{GJ}$,
 (iii) a tail of plasma
distribution with the energy up to $\,  \gamma_t =10^4- 10^5$.

The choice of a particular distribution function (Fig.
\ref{Distributionfunction}) is very important. All the following
is   dependent on this  choice. In  particular, in our model, the primary
beam is composed of electrons or positrons. We show
 in Appendix \ref{Instabilityion} that both
cyclotron-Cherenkov and Cherenkov-drift instabilities do not develop
in  an ion beam.

The {\it electromagnetic}  cyclotron-Cherenkov and Cherenkov-drift instabilities
are  the strongest instabilities
in the pulsar magnetosphere  (\cite{LyutikovInst}).
This differs from the more common case of a nonrelativistic plasma,
where electrostatic Cherenkov-type
 instabilities (i.e. those that result in  emission of electrostatic
Langmuir-type waves) are generally
stronger than electromagnetic instabilities.
In addition, for a one-dimensional plasma
streaming along the  magnetic field, the effective
parallel mass is considerably increased by  relativistic effects. 
For the particles in the primary beam, which contribute to the
development of the instability, the effective parallel mass is 
$m_{eff\parallel} \,= \, \gamma_b^3  m \approx 10^{21}\, m$ ($ m$ is a mass of a
particle). 
This
suppresses the development of the electrostatic instabilities. 
In contrast,
the effective transverse mass , $m_{eff\perp} = \gamma_b  m$,
is less  affected by the large parallel
momentum. The electromagnetic instabilities
are less  suppressed by the large streaming momenta. 
Thus, the relativistic velocities and one-dimensionality
of the distribution function result in a strong suppression
of the electrostatic instabilities and strengthen
electromagnetic instabilities.

Cyclotron-Cherenkov
generation of wave by fast particles is not new in astrophysics.
For example, cosmic
rays in the interstellar medium and in  supernova shocks generate
Alfv\'{e}n by a similar
 mechanism. In the case of Alfv\'{e}n waves in the nonrelativistic
electron-ion plasma,  the frequency of the waves $\omega$ can be much
smaller than the $k v$ term and can be neglected in the
resonance condition. The important difference between these applications
and cyclotron-Cherenkov
instability in pulsar magnetosphere is that the generated waves belong
not to the hydromagnetic
Alfv\'{e}n  waves, that cannot leave  the plasma, but to near vacuum
electromagnetic waves.

We should also mention that a cyclotron-Cherenkov instability of
an electron beam propagating along a  magnetic field is  known
in the laboratory as a very effective  source of the high frequency
microwave radio emission (\cite{Galuzo}, \cite{Didenko},
\cite{Nusinovich}).
The so called slow-wave electron cyclotron masers (ECM) provides  a
high efficiency and  high power microwave source.
Though there are no commercial slow-wave ECM available now, they
 are believed to be
very promising devices due to their better control of the beam
quality and potentially  more compact design than the cyclotron
autoresonance masers.
Thus,  pulsars can be regarded  as cosmic slow-wave ECMs.

In  previous work (\cite{LyutikovMachabeliBlandford1})
we developed a new
approach to the amplification of curvature radiation.
We argue, that a new,
  Cherenkov-drift
instability, may be operational in the pulsar magnetosphere.
 The  Cherenkov-drift emission  combines the features of both Cherenkov
and curvature emission processes.
 This instability is similar to the drift instabilities of the inhomogeneous
 plasma. The striking feature is that, unlike the
 nonrelativistic laboratory plasma, where
 drift instabilities develop on the low frequency waves with the
 wave length on the order of the inhomogeneity size,
 in the strong relativistic plasma  drift instabilities can produce
 high frequency waves.
 We develop an approach that treat
Cherenkov and curvature emission consistently in cylindrical coordinates.
The choice of cylindrical coordinates allows one to consider
curvature emission as a resonant emission process.
(in the former approaches the wave-particles
interaction length was very limited, that precluded a  strong amplification
under all circumstances).
Another important difference in this work is the proper account of the
dispersion and polarization of the normal modes. We show that
 Cherenkov-drift
instability develops only in a medium which supports subluminous waves.

We argue that the  theory  based on the cyclotron-Cherenkov
and Cherenkov-drift instabilities
is capable  of  explaining a very broad range
of the pulsars' observed  properties. 
In a "standard" pulsar with a surface magnetic field $B=10^{12}\,G$
and a period $P=0.5$ s both cyclotron-Cherenkov and 
Cherenkov-drift instabilities occur at a radius of about $10^9 $ cm.
In a dipole geometry
  the emission region is limited to the field
lines near the direction of the  magnetic moment of the neutron star.
The emission region for the Cherenkov-drift instability is larger
than for the cyclotron-Cherenkov instability. In both cases it is 
determined
by the curvature of the magnetic field lines that  limit the coherent
growth of the waves.

In Section \ref{Observational properties} we shortly review the observational
 properties of the pulsar radio emission.
 In Section \ref{instabilities} we discuss the microphysics 
of the underlying cyclotron-Cherenkov and Cherenkov-drift instabilities.
In Section \ref{Model} we describe the fiducial pulsar
model,
and finally in Section \ref{PulsarPhenomenology} we show how the 
 properties of the pulsar
radio emission may be explained in the framework of this theory.

\section{Observational  properties and phenomenological theory of pulsar
radio emission}
\label{Observational properties}

\begin{figure}
\psfig{file=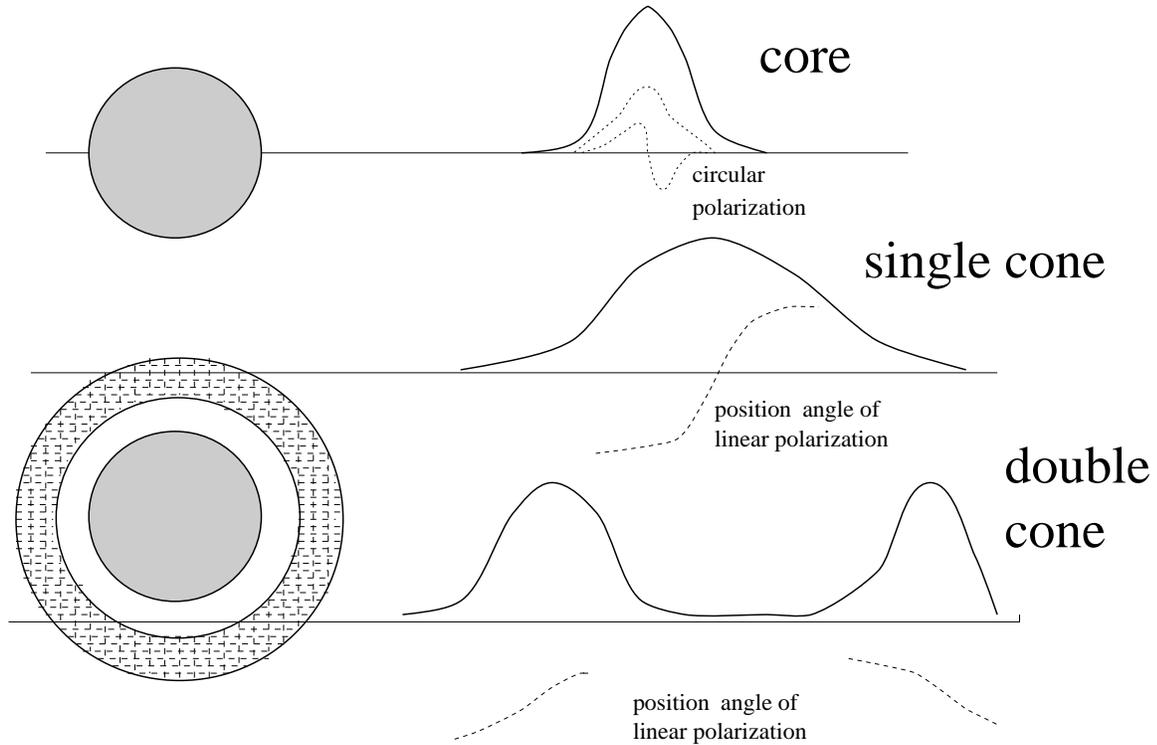,width=15.0cm}
\caption{
Example of emission geometry that produces core and cone profiles.
\label{emissiongeometry}}
\end{figure}

A useful observational framework for discussing theory is 
a description 
of pulsar
radio emission given by \cite{rankin1}. The main feature of this model
is the division of emission  into two main classes: core and cone.
There  may be many cones of emission. In each pulsar
the averaged profile may be a combination of core and/or cone emissions
(Fig. \ref{emissiongeometry}).

The majority of pulsars (about 70{$\%$})
 show core-type emission. 
The typical core emission has the following
features: (a) the profile has a single component,
(b) variable circular polarization (up to 60{$\%$}),
the amount of the circular polarization either reaches maximum
at the maximum intensity or the sense of polarization
may reverse in the middle of the pulse (Fig. {emissiongeometry}),
(c) linear polarization changes
from nearly 100$\%$ to unpolarized, 
in most cases the radiation may be split into two
orthogonally polarized modes (\cite{Stinebring1984}).

About 30{$\%$} of all pulsars show pure conal emission and they are divided
into two main groups -- cone singles and cone doubles which are believed to
be closely connected, the only difference being the geometrical path of the
line of sight through emission region.
Conal type emission shows a great variety of phenomena. Some of the 
 typical features of conal emission are
(a) the profiles can have up to four components, corresponding to two cones,
(b) circular
polarization is small and unsystematic, (c) linear polarization is moderate,
some pulsars show a
sudden change of position angle by $\pm \pi/2$;
in most cases the radiation may be split into two
orthogonally polarized modes, then the change of position 
angle by $\pm \pi/2$ corresponds to the change in the relative intensity
of the modes.
Besides these phenomena,  cone-type emission shows drifting subpulses,
nullings and  mode switching. These effects are probably related to the
temporal and/or spatial  modulations of the parameters of the 
outflowing plasma and will not be discused here (see \cite{KazbegiShukre}).

There have been contradictory  attempts to determine observationally the emission 
altitude but this conclusion is model-dependent.
By contrast,    \cite{Gwinn}  used interstellar scattering to measure  directly
the size of the emission region of 
 $\approx 500$
km ($\approx 0.1 c/\Omega$).  The other, model-dependent,
observational evidences for high emission altitudes 
comes from the large duty cycles  often 
observed in  pulsars ("wide beam pulsars",
e.g., \cite{LyneManchester1988}).
Conventionally, these pulsars are interpreted as almost aligned
rotators.  An alternative explanation is that the emission may be coming
from large radii.

Another class of pulsars, 
those with interpulses, are conventionally interpreted as orthogonal 
rotators with emission coming from two poles despite the fact  that emission bridges  are often observed  in these
pulsars   \cite{HankinsFowler1986} and, at least in one case,  PSR B0950+08 the 
polarization data imply a nearly aligned rotator (\cite{Manchester1995}).
The Crab pulsar also  has a bridge of emission 
between the two  pulses, intensities of the main pulse and interpulse 
are correlated and the angle of linear polarization in the two
pulses seems to be related.
In addition, it shows additional emission peaks {\it between} the interpulse
and pulse  at  centimeter
wave length \cite{Moffet}.
 \cite{Manchester1996} suggests
that   interpulses come from the same pole. If so, the simplest interpretation
is that emission originates at high altitudes.

\section{Cyclotron-Cherenkov and Cherenkov-drift instabilities}
\label{instabilities}

In this section we consider the physics of the 
cyclotron-Cherenkov and Cherenkov-drift instabilities (\cite{Ginzburg},
 \cite{LyutikovMachabeliBlandford1}).
The terminology used here to describe these instabilities refers to the
fact that in the cyclotron-Cherenkov emission,
 a resonant particle changes its gyrational  state  
(undergoes a transition between different Landau level), thereby comes
the "cyclotron" part of its name, but the force that induces
the emission is due to the presence of a medium
(the "Cherenkov" part of the name).
The Cherenkov-drift emission is similar to conventional 
Cherenkov (the gyrational
 state of the resonant particle
remains unchanged)  but it involves a nonvanishing curvature
 drift of the resonant particles.

The interplay between cyclotron
 (or synchrotron) and Cherenkov radiation has been
a long-standing matter of interest.
 \cite{Schwinger} discussed the
relation between  these two seemingly different emission mechanisms.
They showed  that
 conventional synchrotron emission
and Cherenkov radiation may be
 regarded as respectively limiting cases
 of $|n-1|\, \ll \,1 $ and $B=\,0$
of a synergetic (using the terminology of Schwinger  {\it et al.}
\cite{Schwinger}) cyclotron-Cherenkov radiation.
In the work ( \cite{LyutikovMachabeliBlandford1}
this analogy has been extended to include
 inhomogeneity of the magnetic field.

The 
physical origin of the emission in the case of Cherenkov-type and
synchrotron-type
  processes is quite different. In the case of   Cherenkov-type
process, the emission may be  attributed to the electromagnetic
polarization shock front that
develops in a dielectric medium due  to the passage of a charged particle with
speed larger than phase speed of  waves in a medium. It is virtually a collective
emission process.  In the case of
synchrotron-type  process, the emission may be attributed to the Lorentz force acting
on a particle in a magnetic field.  Cherenkov-type
emission is impossible in
 vacuum and in a medium with the refractive index smaller than
unity.

Both the cyclotron-Cherenkov and Cherenkov-drift
instabilities, that we believe can develop in pulsar
magnetosphere, operate in the kinetic regime, i.e. they  are of a maser type
(\cite{LyutikovPhD}).
This means that there is some kind
of  population inversion in the phase space, which
supplies the energy for the development of the instability.
In the present case,
the source of free energy is the
anisotropic distribution function of the fast particles.
The condition of a  population inversion  may be restated that
the induced  emission dominates over induced
absorption for a given transition.

The first two steps in identifying the possible
maser-type radio emission generation
mechanism are (i) determining which radiative  transitions are
allowed in a given system and (ii) establishing if the
given distribution function allows for the population inversion
for the particle in resonance with the emitted waves.
In this chapter we will first discuss the microphysics
of the two suggested emission mechanisms and then show that
the distribution function of the particles present on the open field
lines of pulsar magnetosphere does have a population
inversion and allows  maser action.
When discussing the  microphysics
of the emission process, we will concentrate on the spontaneous  emission
processes. The induced emission rate, which is important for the
development of the instabilities, is
derivable from the spontaneous emission in the usual manner.
In the process of   induced  emission,
 the electron emits a wave in  phase with the incident wave.
However, both   cyclotron-Cherenkov and Cherenkov-drift
masers are broadband and incoherent because a
single electron can resonate with
several waves simultaneously.

\subsection{Physics of Cyclotron-Cherenkov Emission}
\label{cyclotronCherenkov}

The 
cyclotron-Cherenkov instability develops at the anomalous 
cyclotron resonance 
\begin{equation}
\, \omega({\bf k}) - k_{\parallel} v_{\parallel} + {\, \omega_B \over  \gamma } =0
\label{a}
\end{equation}
where $\, \omega({\bf k})$ is the frequency of the normal mode,
${\bf k}$ is a wave vector, $v$ is the velocity of the 
resonant particle, $\, \omega_B= |e| B /mc $ is the nonrelativistic 
gyrofrequency, $\,  \gamma$ is the Lorentz factor  in the
pulsar frame, $e$ is the charge of the resonant particle, $m$ is its
mass and $c$ is the speed of light.   
Note a sign before the $\, \omega_B$ term. 

To describe the microphysics of the cyclotron-Cherenkov emission
emission process we first recall the microphysics 
of the conventional Cherenkov emission (\cite{Ginzburg}).
Consider a charged particle propagating in an unmagnetized 
dielectric with the dielectric constant $\epsilon > 1$.
 As the particle propagates, it induces 
a polarization in a medium. If the velocity of the particle
is larger than the velocity of propagation of the 
polarization disturbances in a medium, which is equal to the 
phase speed of the waves $v_{ph} = c/\sqrt{\epsilon} \, < c $,
the induced polarization cannot keep up with the particle. 
This results in a formation of the polarization shock front.
 At large distances,
 the electromagnetic fields
from  this "shock front" have a wave-like form corresponding to
 Cherenkov emission.
 Thus
the emission is attributed to the polarization shock front and
not directly to the particle. This polarization shock front acts
on the particle with a drag force, which slows down
the particle. This drag force may be considered
as a generalization of the radiation reaction force in a medium.

Now let us consider the propagation of a particle in a magnetized
dielectric along a spiral trajectory.
 Similarly, the propagating charged particle induces
polarization in a medium.
  If the velocity of the particle
is larger than the phase speed of the waves, a polarization
shock front develops, which acts
on the particle with a drag force. Now the drag force, 
averaged over the gyrational period, has two
components:  along the external magnetic field and perpendicular
to it. The parallel part of the drag force always slows
the particle down. The surprising result is that the 
perpendicular component of the drag force acts to {\it increase}
the transverse momentum of the particle. Thus a particle
 undergoes a transition to the state with {\it higher} transverse momentum 
and {\it emits} a photon. The energy is supplied by the parallel motion.

The photons emitted by such mechanism correspond to
 the anomalous Doppler effect
$\, \omega-\, k_{\parallel} v_{\parallel} - s \, \omega_B/\,  \gamma_{res}=0$,
with  $s\, < 0$.  In a vacuum,
 only the normal Doppler resonance, with  $s\, >  0$,  is possible.
The necessary condition for the anomalous Doppler effect,
 $\, \omega -\,  k_{\parallel} v_{\parallel} \,<\, 0$,  may be satisfied for fast
particles propagating in a medium with the refractive index larger than unity.
It is natural to attribute the emission 
at the normal Doppler resonances to the
Lorentz force  of the magnetic field
acting on the electron,
 while  the emission at the anomalous
Doppler resonances is attributed 
 to the electromagnetic drag forces from the medium.

It may also
 be  instructive to consider the emission process in the frame associated
with the center of gyration of the particle. In that frame the
waves emitted at the anomalous Doppler effect have {\it negative} energy, so
that in the emission process the particles increases its energy and
emits a photon.

The  cyclotron-Cherenkov instability may be considered as a
maser using the induced cyclotron-Cherenkov emission.
The free energy for the growth of the instability
comes from the nonequilibrium
anisotropic distribution of fast particles. The condition
that the emission rate dominates the absorption requires population
inversion in the distribution function of fast particle
(maser action). Since radiation
reaction due to the anomalous Doppler effect
 induces transition up in quantum levels , for the instability to
occur we need more particles on the {\it lower} levels. From the  kinetic
point of view, waves grow if the quantity ${\bf k} \,{\partial f({\bf p})
\over \partial {\bf p}}$ is positive for some values of ${\bf k}$. For
an electron in a  magnetic field this condition takes the form
\begin{equation}
{s \, \omega_b\over v_{\perp}} { \partial f\over \partial p_{\perp}} + 
k_{\parallel}
{\partial f\over  \partial p_{||}} \, > 0
\label{ffwq}
\end{equation}
where 
$s$ is a harmonic number.
Here $s>0$ corresponds to normal
Doppler effect (transition down in Landau levels) and $s<0$ corresponds
to anomalous Doppler effect (transition up in Landau levels). If the
distribution function is a plateau-like in parallel momenta then the
condition for  instability is $ s {\partial f\over \partial p_{\perp}} >0$
which could be satisfied for inverted population for the normal
Doppler effect or for the "direct" distribution for anomalous Doppler
effect. The latter case takes place for the beam of particles propagating along
the
magnetic field with no dispersion in transverse momenta.

\subsection{Physics of Cherenkov-Drift Emission}
\label{Cherenkovdrift}

There is  a possibility for the development  of the 
Cherenkov-drift instability, which occurs at the resonance
\begin{equation}
\, \omega({\bf k}) - k_{\parallel} v_{\parallel} - k_{\perp} u_d  =0
\label{a1}
\end{equation}
where $u_d = \,  \gamma v_{\parallel} c / \, \omega_B R_B$ is the drift 
velocity, $ R_B$ is the curvature radius of the magnetic field line.
 A weak inhomogeneity of the magnetic field
results in a curvature  drift motion of the particle perpendicular to the
local plane of the magnetic field line. A gradient drift
(proportional to $ ( {\bf   B \cdot \nabla } )  {\bf B} $
much smaller than the curvature drift and will be neglected.
When the motion of the particle
parallel to the magnetic field is ultrarelativistic,
the drift motion  can become weakly relativistic 
even in a
weakly inhomogeneous field resulting in the
 generation of  electromagnetic, vacuumlike waves.
 The presence of three ingredients  (strong but finite magnetic
field,  inhomogeneity of the field  and a medium with the index of refraction larger
than unity) is essential for this type of
 emission.  We will call this mechanism
Cherenkov-drift emission stressing the fact that microphysically it is virtually
Cherenkov-type emission process.

Conventional consideration of the curvature emission (\cite{Blandford1975},
\cite{ZheleznyakovShaposhnikov},
\cite{MelroseLou},  \cite{Melrosebook1}) emphasize the
analogy between  curvature emission and  conventional cyclotron emission.
To our opinion this approach, though formally correct, has  limited
applicability and  misses some  important
physical properties of the emission mechanism. 
It has two important shortcomings.
  The first is that in adopting
a plane wave formalism, the interaction length for an individual
electron,  $\approx R_B /\gamma_b$,
is essentially coextensive with  the region over which the waves 
interact with a single
 electron. The approach necessarily precludes   strong amplification
under all circumstances because the wave would have to grow substantially
during a single interaction in a manner that could not be easily
quantified. The second problem was a neglect of dispersion.
We address the first shortcoming by expanding the electromagnetic
field in cylindrical waves centered on $r=0$, and the second
explicitly by considering general plasma modes.

 In a separate approach Kazbegi {\it et al.}
\cite{Kazbegi} considered this process calculating a dielectric
tensor of an inhomogeneous magnetized medium, thus treating the
emission process as a collective effect. They showed that  maser
action is possible only if a  medium supports subluminous waves.
In the previous work (\cite{LyutikovMachabeliBlandford1}
 we showed  how these two approaches can be reconciled and
argue that the  dielectric
tensor  approach, which treats the Cherenkov-drift emission as a collective
process, has a wider applicability.

It is more natural to consider  Cherenkov-drift emission in a curved
magnetic field as an analog of the Cherenkov  emission with the
drift of the resonance particles taken into
account,  than as the type of a curvature
emission. From the microphysical point of view
 the emission is again due to the
polarization shock front that develops due to the passage of a superluminal
particle through a medium, so it  is required
that the emitting particle propagates with the
velocity greater than the phase velocity of the emitted waves. The
 Cherenkov-drift maser is impossible in vacuum, unlike the  curvature
emission, which is a close analog of the conventional cyclotron
emission and is possible in vacuum.
The curvature provides only the drift component of the velocity, which
is essential for the coupling  the  resonant particle to the emitted
electromagnetic wave.

In a Cherenkov-type  emission
the resonant particle can  interact  only with the
part of the electric field parallel to the velocity. Thus, if the drift
of the resonant particles  perpendicular to the plane of the field line
is  taken into account,
 it becomes
 possible to
emit a {\it  transverse electromagnetic}
 wave with the electric field along the drift velocity, i.e.
perpendicular to the plane of the curved field line (see Fig.
\ref{Cherenkov-driftemission}).
 The growth occurs on the rising part of the
parallel distribution function where ${\partial f_{\parallel} \over
\partial p_{\parallel} }> 0$. This is satisfied for the particles of the
primary beam.

\section{Model of the Pulsar Radio Emission }
\label{Model}

\subsection{ Plasma generation}
\label{Plasmageneration}

Rotating, strongly magnetized
 neutron stars induce strong electric fields that pull
the charges from their surfaces. Inside the closed  field
lines of the neutron star magnetosphere, a steady charge
distribution established,  compensating the induced electric
field. On the open field lines,  the neutron star
 generates a dense flow of relativistic
electron-positron pairs penetrated by a highly relativistic
 electron or positron beam.
The density of the primary beam is roughly
equal to the Goldreich-Julian density $n_{GJ}={\bf \Omega \cdot B}/( 2\,
\pi\, e\, c)$.
We will normalize the  density of the pair plasma  to the
 Goldreich-Julian density.
\begin{equation}
n_p
= \lambda\, n_{GJ} =10^3-10^6 n_{GJ},
\hskip .3 truein
\omega_p^2 = \lambda \omega_b^2 = 2  \lambda \omega_B \Omega
\label{qpq}
\end{equation}
where $\lambda$ is the multiplicity
factor which is the number of pairs produced by each primary particle.
Secondary pairs are born with almost the same energy
 in the avalanche-like
process above in the polar cap  (\cite{Arons83}).
 The pair creation front in the polar cap region
is expected to be very thin so we can in the first approximation neglect
the  residual electric field in the front that could lead to the
reversed current and different initial
energies and densities of secondary particles.
The combination of the pair plasma and primary beam is expected to screen
the rotationally induced electric field so that the flow is force-free.

Another relation between the parameters of the plasma and the beam
comes from the energy argument that the primary particles stop producing
the pairs when the energy in the pair plasma becomes equal to the energy in the
primary beam:
\begin{equation}
2 \,
<\,\gamma\,>^{(0)}_{\pm}\, \lambda \approx \gamma_b , \hskip .6 truein
\mbox{at the pair formation front}
\label{vg}
\end{equation}
where it was assumed that the initial densities, temperatures
 and velocities of the plasma
components are equal. For cold components $<\,\gamma\,>^{(0)}_{\pm}\,=\,
\gamma_p$ while for the relativistic  components with a temperature
$T_p$ the average energy is
 $<\,\gamma\,>^{(0)}_{\pm}\,=\, 2\, \gamma_p \, T_p$.
The assumption of equipartition (\ref{vg}) is very approximate one, but
it allows considerable simplification of numerical  estimates.
If for some reason, this would turn out to be an incorrect assumption,
the corresponding formula can be adjusted by changing the scaling.

As an estimate of the densities of the particles from the
tail of plasma distribution we will use the assumption that
the energy in the tail is approximately equal to the energy in the
plasma  (and in the beam):
\begin{equation}
\, \gamma_t n_t  \approx  \, \gamma_b^{(0)}
\label{v2}
\end{equation}
where $\, \gamma_t$ and $n_t$ are the typical energy and the density
of the tail particles.

\subsection{Typical pulsar}
\label{typical}

In this work we will make numerical estimates for the "typical" pulsar
with the following parameters:

(i) magnetic field
is assumed to be dipole with the magnetic field strength at the surface of neutron
star $B_{NS} =\, 10^{12} G$,

(ii) rotational period of the star $P= 0.5 \,$ s (light
cylinder radius $R_{LS}= 2.4 \times 10^9$ cm).

(iii) average streaming energy of the plasma components $\gamma_p=10 $

(iv) temperature of the plasma components $T_p =10$.

(v) initial energy of the primary beam  at the pair formation front
$\gamma_b =6 \times 10^7$.

The  energy of the beam will decrease due
to the curvature radiation reaction force
(Appendix \ref{cooling}).
 Then at the light cylinder,
where the instabilities occur, the beam will decelerate due to the curvature
radiation reaction to $\,  \gamma_b = 2 \times 10^6$.

For a given period and magnetic field the
equation (\ref{vg}) reduces the number of free parameters
for the plasma to two: the
plasma temperature and the bulk streaming energy $\,  \gamma_p$
(or temperature and the multiplicity factor $\, \lambda$).
We chose a strongly relativistic plasma with the invariant temperature
$T_p =10$.
The multiplicity factor $\lambda$ corresponding to these parameters follows from
Eq. (\ref{vg}): $\lambda = 3 \times 10^5$.
 The average energy of the tail particles is assumed to be
$\, \gamma_t =10^5$.
An important factor that determines the growth rate of the
instabilities is the energy scatter of the resonant particles.
In  estimating   the growth rates of the
cyclotron-Cherenkov and Cherenkov-drift instabilities on the primary beam
 we will also
assume
that the scatter in Lorentz factors of the primary particles in the
pulsar frame $\, \Delta \gamma =  10^2$.
\label{Deltagamma}
This assumes that the beam has cooled considerably due to the
curvature radiation and lost about 90\% percent of its energy.

Several
 points are important in our choice of parameters.
First, we use a
 relatively low plasma streaming $\,  \gamma$-factor (and respectively
high multiplicity factor $\, \lambda$).
In the polar cap models (\cite{Arons81a}, \cite{Arons83})
the pulsar plasma will have a low streaming $\,  \gamma$-factor if the
the magnetic field near the surface differs considerably from the dipole
field thus reducing the radius of curvature
(\cite{MachabeliUsov1989}). Secondly, the required scatter in energy
of the primary beam particles ($\Delta \gamma =  10^2$) is very
small. This is due to the effects of curvature radiation reaction
on the primary beam during its propagation through the dipole
pulsar magnetosphere. The highly nonlinear damping rate due the
emission of curvature radiation by the primary beam result in an
effective cooling of the beam (see Appendix \ref{cooling}).

We used two types of the particle distribution function to calculate the
relevant moments: water bag and relativistic Maxwellian (\cite{LyutikovPhD}).
For the case of relativistic Maxwellian distribution function the
strongly relativistic temperature $T_p =10$
 implies that, formally, there are many backward
streaming particles. We note, though, that the backward
streaming particles  do not contribute significantly to any of the relevant moments
of the distribution function, so that we can regard the strongly relativistic
streaming Maxwellian distribution as a fair approximation to the relatively unknown, but
definitely very hot,
real distribution function.

\begin{figure}
\psfig{file=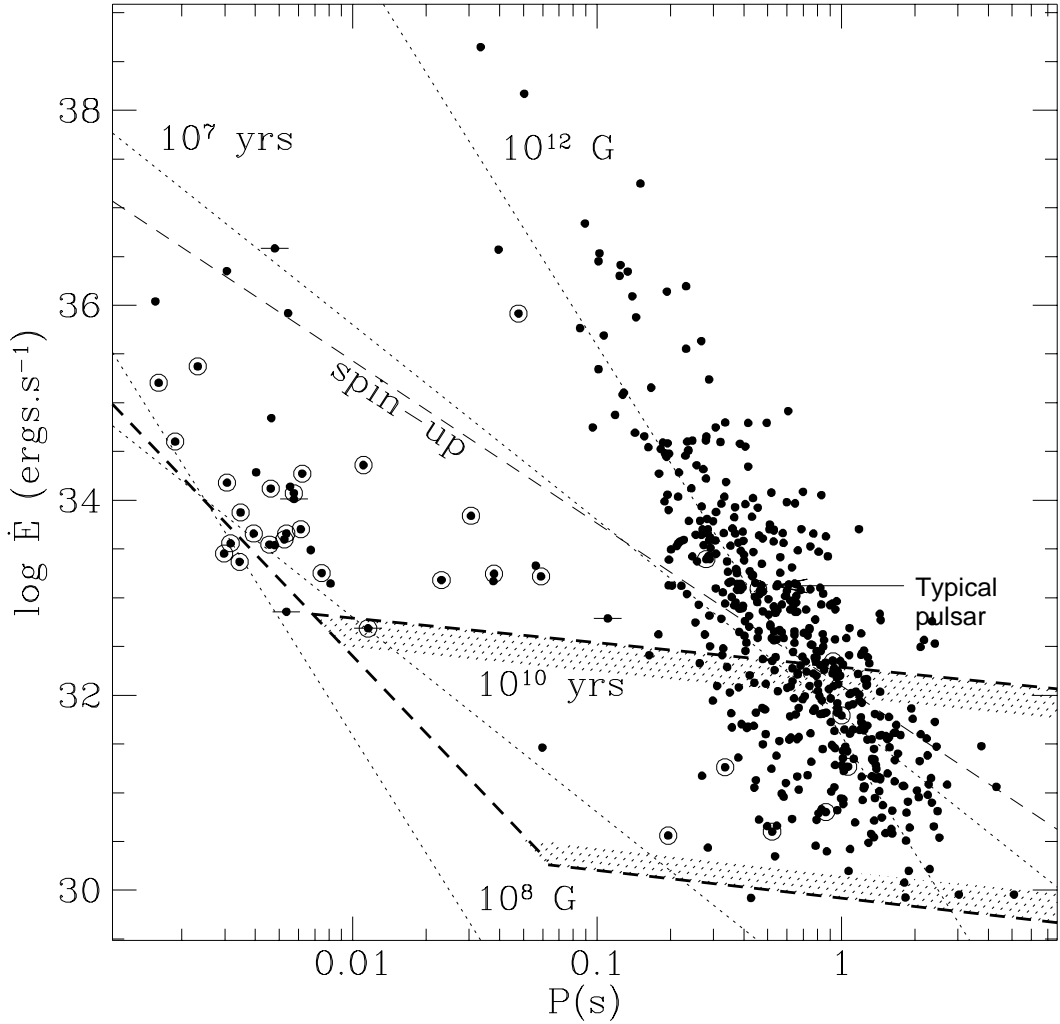,width=15cm}
\caption[P-$\dot{E}$ diagram]{ The dotted lines show constant spin down
ages and magnetic fields. There is also a dashed line for the
equilibrium period spin-up. The heavy dashed curves and shading illustrate
several death lines from the literature (see \cite{Hansen} for references).
The circles indicate binary pulsars and
those points with horizontal lines through
them are in globular clusters with apparently negative
$\dot{E}$ (they are
contaminated by cluster accelerations).
}
\label{PPdot}
\end{figure}

\begin{figure}
\psfig{file=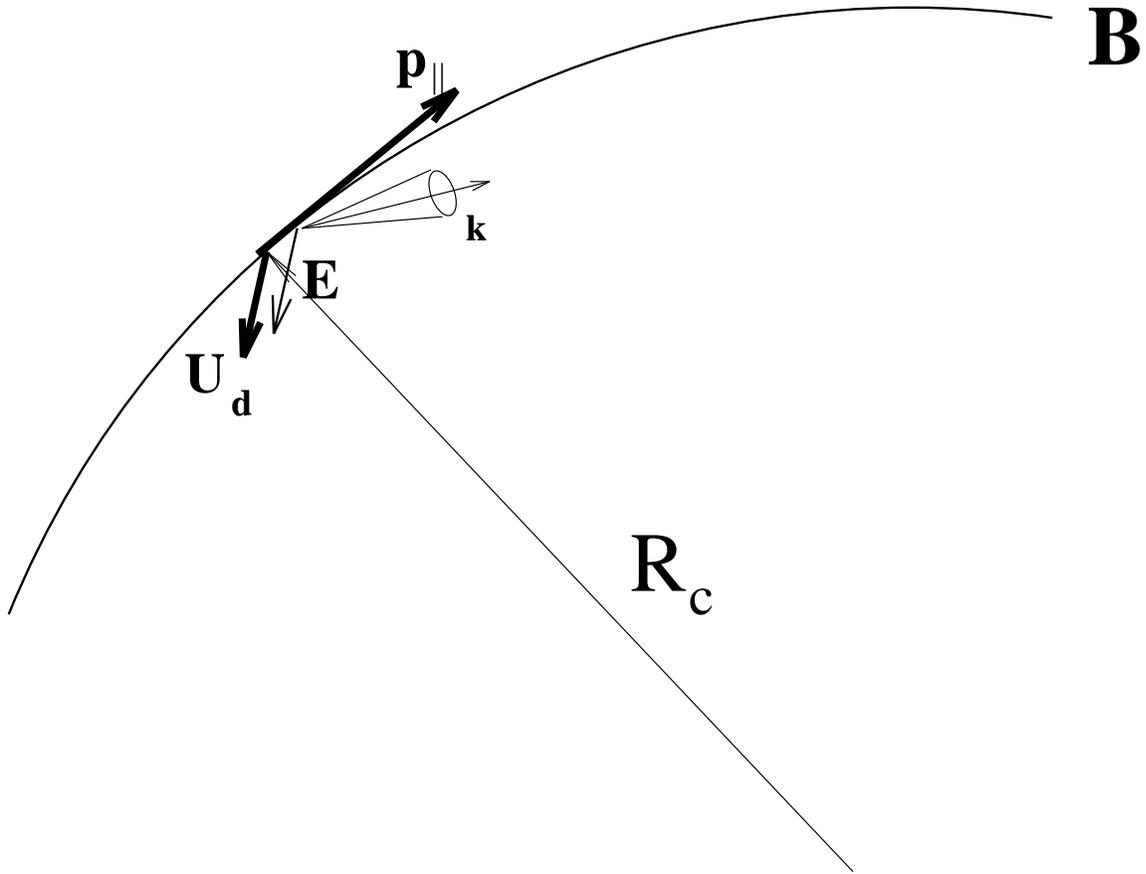,width=15.0cm}
\caption{
Cherenkov-drift emission. Drift velocity ${\bf u}_d$ is perpedicular
to the plane of the curved field line $({\bf B}-{\bf R_c}) $
plane, ${\bf R_c} $ is a local radius of curvature).
The emitted electromagnetic waves
 are polarized along  ${\bf u}_d$. The emission is generated in the
 cone centered at the angle $\theta^{em} = u_d / c $ and
 with the opening angle  $(2 \delta)^{1/2} \ll \theta^{em}$.
\label{Cherenkov-driftemission}
}
\end{figure}

The location of our typical pulsar on the $P - \dot{E}$ diagram is shown
on Fig \ref{PPdot}.
The corresponding plasma densities and frequencies are given in Tables
\ref{Thesesta1}  and \ref{Table2} for the
two locations (near the stellar surface and at the emission region $ R \approx
10^9$ cm in the pulsar and plasma reference frames.

\begin{sidewaystable}
\centering
\begin{tabular}{|l|c|c|}
\hline
   & Pulsar frame
 & Plasma frame \\ [0.3cm] \hline
Magnetic field, G & $10^{12}$ &$10^{12}$ \\ [0.3cm]  \hline
Cyclotron frequency ${\omega_B}^{\ast}$, $( {\rm rad\,s^{-1}})$,
& $1.8\times 10^{19}$ & $1.8\times 10^{19}$ \\ [0.3cm] \hline
Beam density, $cm ^{-3}$  & $ n_{GJ}^{\prime}\,=\,
\Omega B /( 2\, \pi\, e\, c) =
1.4\times 10^{11}$ &  $ n_{GJ}\,=\,
 \Omega B /( 2\, \pi\, e\, c \gamma_p) = 1.4\times 10^{10}
$ \\ [0.3cm] \hline
Beam plasma frequency,  $ {\rm rad\, s^{-1}}$ & $
\omega_{GJ}^{\prime}\,=\, \sqrt{ { 4\, \pi\, q^2 \, n_{GJ}^2 \over m}} \,=\,
\sqrt{ 2 \Omega \omega_B} \,=\,
2\times
 10^{10}$ & $ \omega_{GJ}\,=\, { \omega_{GJ}^{\prime} \over \sqrt{ \gamma_p}
 }\,=\,\sqrt{{ 2 \Omega \omega_B\over \gamma_p} } \,=\,
6.3\times
 10^{9} $ \\ [0.3cm] \hline
Beam energy &  $\gamma_b^{\prime} \,=\, 6\times 10^7$ &
$\gamma_b \,=\, { \gamma_b^{\prime} \over 2 \gamma_p} \,=\,3 \times 10^6$
\\ [0.3cm] \hline
Plasma density, $cm ^{-3}$  & $ n_p^{\prime}\,=\,{\lambda \,
\Omega\, B \over  2\, \pi\, e\, c} =
4\times 10^{16}$ &  $ n_p,=\,{\lambda \,
 \Omega B \over 2\, \pi\, e\, c \gamma_p}= 4\times 10^{15}
$ \\ [0.3cm] \hline
Plasma frequency, ${\rm rad\, s^{-1}}$
 & $ \omega_p^{\prime} \,=\,\sqrt{ 2 \,{\omega_B} \, \Omega \,
\lambda} = 1.2\times 10^{13}$ &
$\omega_p \,=\,\sqrt{ { 2 \,{\omega_B} \, \Omega \,
\lambda\over \gamma_p}} \,=\, 4\times 10^{12} $ \\ [0.3cm] \hline
$1-\beta_{X}$, $\beta_X$- phase speed of X mode &
$\phantom{{{{{a\over b}\over{a\over b}}}\over{{{a\over b}\over{a\over b}}}}}
 \delta^{\prime}
={ \omega_p^{\prime \,2} T_p \over 4 \gamma_p^3 \omega_B^2} =
{ \lambda  \Omega T_p \over 2 \gamma_p^3 \omega_B } = 10^{-15}   $&
$\delta = { \omega_p^2 T_p \over \gamma_p \omega_B^2} =
{ 2  \lambda  \Omega T_p \over \gamma_p \omega_B } =
 4\times  10^{-13}  $ \\ [0.3cm] \hline
\end{tabular}
\caption[Plasma parameters at the surface of the neutron star]
{Plasma parameters at the surface of the neutron star}
\label{Thesesta1}
\end{sidewaystable}

\begin{sidewaystable}
\centering
\begin{tabular}{|l|c|c|}
\hline
   & Pulsar frame
 & Plasma frame \\ [0.3cm] \hline
Magnetic field, G & $10^{3}$ &$10^{3}$ \\ [0.3cm] \hline
Cyclotron frequency ${\omega_B}^{\ast}$, $({\rm rad\,s^{-1}})$,
& $1.8\times 10^{10}$ & $1.8\times 10^{10}$ \\ [0.3cm] \hline
Beam density, $cm ^{-3}$  & $ n_{GJ}^{\prime}\,=\,
\Omega B /( 2\, \pi\, e\, c) =
1.4\times 10^{2}$ &  $ n_{GJ}\,=\,
 \Omega B /( 2\, \pi\, e\, c \gamma_p) = 14
$ \\ [0.3cm] \hline
Beam plasma frequency,  ${\rm rad\, s^{-1}}$ & $
\omega_{GJ}^{\prime}\,=\, \sqrt{ { 4\, \pi\, q^2 \, n_{GJ}^2 \over m}} \,=\,
6\times
 10^{5}$ & $ \omega_{GJ}\,=\, { \omega_{GJ}^{\prime} \over \sqrt{ \gamma_p}
 }\,=\,3\times
 10^{5} $ \\ [0.3cm] \hline
Beam energy &  $\gamma_b^{\prime} \,\approx \, \times 10^6$ &
$\gamma_b \,=\, { \gamma_b^{\prime} \over 2 \gamma_p} \,=\,5 \times 10^4$ \\ [0.3cm] \hline
Plasma density, $cm ^{-3}$  & $ n_p^{\prime}\,=\,{\lambda \,
\Omega\, B \over  2\, \pi\, e\, c} =
4\times 10^{7}$ &  $ n_p,=\,{\lambda \,
 \Omega B \over 2\, \pi\, e\, c \gamma_p}= 4\times 10^{6}
$ \\ [0.3cm] \hline
Plasma frequency, ${\rm rad\, s^{-1}}$
 & $ \omega_p^{\prime} \,=\,\sqrt{ 2 \,{\omega_B} \, \Omega \,
\lambda} = 4\times 10^{8}$ &
$\omega_p \,=\,\sqrt{ { 2 \,{\omega_B} \, \Omega \,
\lambda\over \gamma_p}} \,=\, 1.2\times 10^{8} $ \\ [0.3cm] \hline
$1-\beta_{X}$,$\beta_{X}$ phase speed of X mode & $\delta^{\prime}
={ \omega_p^2 T_p \over 4 \gamma_p^3 \omega_B^2} =
{ \lambda  \Omega T_p \over 2 \gamma_p^3 \omega_B } =  5 \times 10^{-7}  $&
$\delta = { \omega_p^2 T_p \over \gamma_p \omega_B^2} =
{ 2  \lambda  \Omega T_p \over \gamma_p \omega_B } = 1.2 \times 10^{-4}
  $ \\ [0.3cm] \hline
Typical frequency, ${\rm rad\, s^{-1}}$ 
& $ \omega^{\prime} \,=\, 5 \times 10^9$ &
$ \omega = { \omega^{\prime} \over 2 \gamma_p} = 2.5 \times 10^8$ \\ [0.3cm] \hline
\end{tabular}
\caption[Plasma parameters at the emission region $R\approx 10^9 $ cm]
{Plasma parameters at the emission region $R\approx 10^9 $ cm}
\label{Table2}
\end{sidewaystable}

The radial dependence of the parameters is assumed to follow
 the dipole geometry of the
magnetic field:
\begin{eqnarray}
&&
{\omega_B}(r)\,=\, {\omega_B}(R_{NS}) \left( { R_{NS}\over r} \right)^3,
\mbox{} \nonumber \\ \mbox{}
&&
\omega_p(r)\,=\, \omega_p(R_{NS}) \left( { R_{NS}\over r} \right)^{3/2}.
\label{vg2122}
\end{eqnarray}
This may not be a good approximation in the outer regions of the
pulsar magnetosphere, where  relativistic retardation,
 currents flowing in the magnetosphere
and the effects of plasma loading  considerably 
change the structure of the magnetic field.

\subsection{Fundamental plasma modes}
\label{plasmodes}

\begin{figure}
\psfig{file=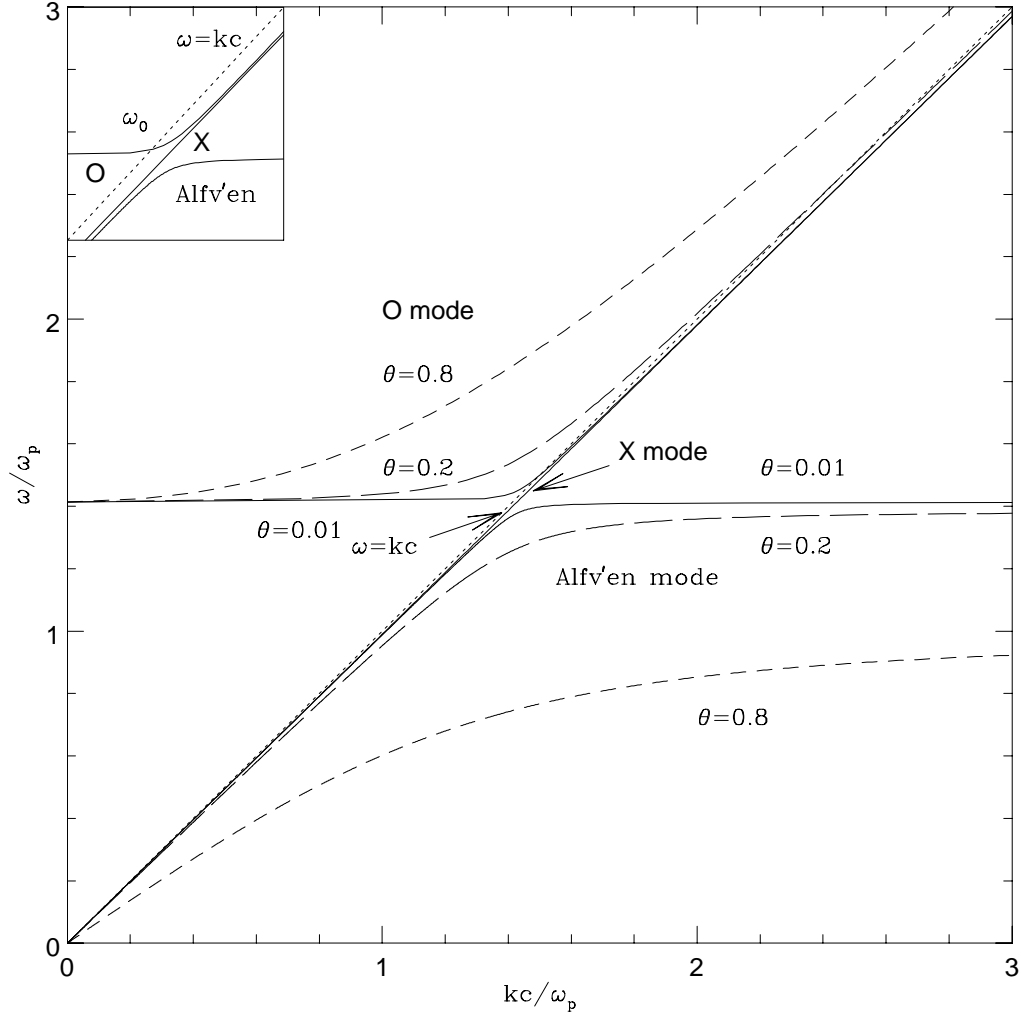,width=15.0cm}
\caption{
 Dispersion curves for the waves in a cold electron-positron
plasma in the plasma frame in the limit $\omega_p \ll \omega_B$. There are three modes
represented by the  dashed (ordinary mode), solid
(extraordinary mode) and long dashed (Alfv\'{e}n mode). The dotted line
represents the vacuum dispersion relation.
 For the exact parallel
propagation,  the dispersion curves for the ordinary mode and
Alfv\'{e}n mode intersect.
The insert in the upper left corner shows the region near the cross-over
point $\omega_0$.
\label{fig1}}
\end{figure}

It is important for the development of both instabilities that the
plasma supported subluminous waves. Such wave indeed exist in a strongly
magnetize electron-positron plasma.
If we neglect (as a first order approximation) the relative streaming of
the plasma  electrons and positrons and the field curvature
then there are three fundamental
 modes: transverse extraordinary wave with the electric field perpendicular
to the {\bf k-B}
plane and longitudinal-transverse mode the electric field in the {\bf k-B}
 with two branches: ordinary
 mode and Alfv\'{e}n mode (\cite{arons1}, \cite{Volokitin}, 
\cite{LyutikovPhD}). For small angles of propagation
with respect to magnetic field
the ordinary mode (upper longitudinal-transverse wave) is quasi-longitudinal
for small wave vector ($k \ll \, \omega_p /c$, $\, \omega_p$ - plasma frequency)
and quasi-transverse for large wave vector ($k \gg  \, \omega_p /c$), while
the Alfv\'{e}n mode is quasi-transverse for small wave vector and
quasi-longitudinal  for large ones.
Fig. \ref{fig1} illustrates these modes in the case of cold plasma components
in the plasma frame and in the low frequency limit $\omega^{\prime} \ll \omega_B$
($\omega^{\prime}$ is the wave frequency in the plasma frame).

The  dispersion relations for the extraordinary mode
in the pulsar frame in the limit $\, \omega \ll \, \omega_B$ is
\begin{equation}
\, \omega = k c  (1- \delta), \hskip .1 truein
\delta = {\, \omega_p^2 T_p \over 4  \, \omega_B^2 \,  \gamma_p^3}  
\label{det3}
\end{equation}
where we used a relationship between the plasma density and the
Goldreich-Julian density $n_p = \, \lambda n_{GJ} $ and the definition
of the Goldreich-Julian density.

For the coupled  Alfv\'{e}n and ordinary modes
it is possible to obtain
the asymptotic expansion
of the dispersion relations
in the limits of very small and very large wave vectors. 
Here we will give the dispersion relations for the quasitransverse
parts of the waves
\begin{eqnarray}
\, \omega^2\,=\, &  {c^2}\,{k^2}\,\left(1 - { \, \omega_p^2 \,T_p\,
\over 2 \,  \gamma_p^3  \, \omega_B^2 }    \,
+ {\frac{2\,  \, \gamma_p{{\, \omega_p}^2} T_p\,{{\sin^2\theta}}}
        {{c^2}\,{k^2}\,
           }} \right) 
\mbox{ O-wave} &
\mbox{ if $ kc \,\gg \sqrt{ T_p} \,\gamma_p \omega_p $ and $\theta < 1/T_p$}
\mbox{} \nonumber \\ \mbox{}
\, \omega^2\,=\, &
 {c^2}\,{k^2}\,\cos^2\theta\,
    \left(1 - { \, \omega_p^2\,T_p\,
\over 2  \,  \gamma_p^3  \, \omega_B^2 }  \,
   - {\frac{{c^2}\,{k^2}\,{{\sin^2\theta}}}
        {2\,T_p\,  \, \gamma_p {{\, \omega_p}^2}}} \right)
\mbox{ Alfv\'{e}n wave}
 &
\mbox{ if $ kc \,\ll\,\sqrt{T_p }\, \gamma_p  \omega_p $}
\label{liafs}
\end{eqnarray}

These relationships are valid for the frequencies satisfying
the inequality 
\begin{equation}
\, \omega \ll \,  \gamma_p \, \omega_B /T_p
\label{x}
\end{equation}
This  is a condition 
 that in the reference frame of the plasma the
frequency of the waves is much smaller that the typical
cyclotron frequency of the particles $\, \omega_B /T_p$.

When the relative streaming of the plasma is taken into account, the
dispersion relations change considerably: a new, slow Alfv\'{e}n branch
appears and the extraordinary and Alfv\'{e}n modes behave differently in the region
of small wave vectors  (\cite{LyutikovDisp}).
The polarization of the fundamental modes also changes.
For the angles of propagation with respect to magnetic field less than some
critical angle, which depends on the difference
of the velocities
of the components and on the magnetic field,
 the two quasi--transverse waves are  circular-polarized,
while for the larger angles  the
two transverse modes become linearly-polarized.

\subsection{Development of instabilities}
\label{Modelemission}

In this section it is shown, that the  pulsar radiation may be
generated  by two kinds of
{\it electromagnetic}  plasma instabilities --
cyclotron-Cherenkov and Cherenkov-drift instabilities. The
cyclotron-Cherenkov instability is responsible for the
generation of the core-type emission and the
Cherenkov-drift instability is responsible for the
generation of the cone-type emission (\cite{rankin1}).
The waves generated by these instabilities
are  vacuum-like electromagnetic waves:
they may leave the magnetosphere
directly.

We assume that the distribution function, displayed in 
Fig. \ref{Distributionfunction}, 
 remains
unchanged throughout the inner magnetosphere. 
This requires that no
Cherenkov-type  two-stream instabilities  develop and that the high
energy  particles
are not excited to the high gyrational states by the mutual collisions
or by the inverse Compton effect.
Then the outer regions  of the pulsar magnetosphere two instabilities can 
develop: (i) cyclotron-Cherenkov instability and (ii) 
Cherenkov-drift instability.

A detailed consideration of the conditions necessary for the 
development of the  cyclotron-Cherenkov  and Cherenkov-drift instabilities
are given in Appendix \ref{Conditioncyclotron} and \ref{Conditiondrift}.
Both cyclotron-Cherenkov  and Cherenkov-drift instabilities
develop in the outer regions of the pulsar magnetosphere at radii
$R \approx 10^9$ cm.

The frequencies of the waves generated by the cyclotron-Cherenkov
instability are given by  (\ref{g1})
\begin{equation}
\omega  = { 4 \,  \gamma_p^3  \omega_B^3  \over
 \, \gamma_{res} T_p \omega_p^2 } 
\label{djjj}
\end{equation}
which may be solved for the radius at which the waves with frequency
$\omega$ are emitted:
\begin{equation}
R =  R_{NS} \left(  { 2 \gamma_p^3 \over \lambda \gamma_{res}
 T_p } \right)^{1/6} 
\, \left(  { \omega_B^{\ast \, 2} \over \omega \Omega  } \right)^{1/6}=
1.8 \times 10^9  \, B_{12}^{1/3} \, \nu_9^{-1/6} \,
P_{0.5} ^{1/6} \gamma_{res, 5} ^{-1/6}  {\rm cm}
\label{djjj1}
\end{equation}
The  relationship (\ref{djjj1})
 may be regarded as a  "radius-to-frequency" mapping. 
For a given $ \,  \gamma_{res}$ the radial dependence of the right hand
side of equation (\ref{djjj}) will result in a radial dependence of the
emitted frequency. The radial dependence
of the parameters  in (\ref{g1}) will  result in emission of higher frequencies
deeper in the pulsar magnetosphere, exactly what is observed.
The frequencies emitted   at the Cherenkov-drift resonance
do not have a simple dependence on radius from the neutron star.
They are determined by several  emission conditions which
limit the development of the instabilities to the  particular
location in the magnetosphere and particular frequencies.

The conditions that the instabilities should satisfy are:
\label{conditions}

(i) small growth length in a curved field lines 
${c / \Gamma } < \Delta \theta R_B$, where $\Delta \theta$ 
is the  range of the emitted resonant angles \label{gr}

(ii) condition of kinetic instability 
$\left| {\bf k} \delta {\bf v}_{res} \right| \gg \Gamma$

In addition to these, the Cherenkov-drift instability
should also satisfy another condition:

(iii) the condition of a large drift
$u_d/c > \sqrt{ 2\delta}$.

The condition (i) states that an emitting  particle can stay in a resonances
with the wave for many growth lengths.
The condition (ii) is a requirement  that the growth rate of 
instability  is much smaller
than the bandwidth of the growing waves.
This condition is necessary for the random phase approximation to
the wave-particle interaction to  apply.

In Appendices \ref{Conditioncyclotron}, \ref{Conditiondrift}
and \ref{Conditionmillisecond} we show that the above conditions can be 
satisfied for the chosen set of parameters 
both in  normal pulsar and millisecond pulsars. 
The conditions for the development of the cyclotron-Cherenkov and
Cherenkov-drift instabilities depend in a different ways on the 
parameters of the plasma. They may develop in the 
different regions of the pulsar magnetosphere. 

\section{Pulsar Phenomenology}
\label{PulsarPhenomenology}

\subsection{Energetics}
\label{Energetics}

In both cyclotron-Cherenkov and Cherenkov-drift mechanism,
the emission is generated by the fast particles which supply the
energy for the growth of the waves.
The total energy available for the conversion into radio emission
is of the order of the energy  of the particle flow along  
the open field lines of pulsar magnetosphere:
\begin{equation}
E\approx   n_{GJ} \, \pi  R_{pc}^2 \,  \gamma_b m c^3 \approx  10^{33} 
{\rm erg \, s^{-1} } 
\label{j}
\end{equation}
where $R_{pc}= R_{NS} \sqrt{ {  R_{NS} \Omega \over c}} $ is the radius
of a polar cap. This is sufficient  to explain the radio luminosities
of the typical pulsar if the effective
emitting area is about   one hundredth of the
total open field line cross section.

\subsection{Emission Pattern}

 The emission  pattern for the "core"-type pulsars
(according to the classification of  (\cite{rankin1})
 is a circle with the angular extent of
several degrees. In our model 
the region of the cyclotron-Cherenkov instability is limited 
to the nearly straight field lines (see Fig. \ref{Deutsch}).

The curvature of the magnetic field destroys the coherence between
the waves and the resonant particles. 
To produce an observable emission the waves need to travel in resonance with the
particle  at least several instability  growth
lengths. Far from the magnetic axis, where the curvature is substantial,
 the waves leave the resonance cone before
they travel a growth length and no substantial amplification occurs.
Near the magnetic field  axis the radius of curvature is very large and
waves can stay in a resonance with a particles for a long time 
and grow to large amplitudes.

By contrast,
the Cherenkov-drift instability requires  curvature
of the field lines,  but its growth rate may be limited
by the coherence condition. In a dipole geometry,
 this will limit  
the Cherenkov-drift emission region to a ring  around the
central field line.
Another possible location of the Cherenkov-drift instability is the
region of the swept back field lines  (Fig. \ref{Deutsch})

The  development of the cyclotron-Cherenkov
instability depends on fewer parameters  than  the 
Cherenkov-drift instability. 
It   develops  on the nearly straight central
field lines. The conditions for the development of the
cyclotron-Cherenkov
instability  (Eqs. \ref{g2},  \ref{ab}, \ref{g10}) depend  on low powers
of the plasma parameters, so it is quite robust.  By contrast,  the
Cherenkov-drift instability depends
on the plasma parameters  {\it and} 
 the radius of curvature of the magnetic field
in a complicated way. This results in a broader range
of phenomena observed in the cone emission.
 If the parameters of the plasma change due the 
changing conditions at the pair production front, the location
of the Cherenkov-drift instability may change considerably. This 
may account for the mode switching observed in the cone emission of 
some pulsars.

It may  also possible  beto explain in the framework of our model the
phenomenon of  "wide beam geometry" observed in some pulsars
(\cite{Manchester1996}). The  Cherenkov-drift instability
may occur in the region, where the field lines are
swept back considerably. Then the emission will be generated in
what could be called a "wide beam geometry".

\subsection{Polarization}
\label{Pola}

If the average energy of the electrons and positrons of the
secondary plasma is the same, the fundamental modes of such strongly
magnetized plasma are linearly polarized.  Both of the two
quasi--transverse modes (one with electric field lying
 in the ${\bf k - B_o}$ plane,
another with $ {\bf E}$ perpendicular to this plan)
 may be emitted by the cyclotron-Cherenkov
mechanism. This may naturally explain the two orthogonal modes
observed in pulsars   (\cite{Kazbegi}, 
 \cite{Machabeli2},  \cite{Machabeli3}). 
The   difference in the dispersion relations
and  in the emission and absorption conditions 
between ordinary and extraordinary modes
 may explain the difference in the observed intensities of these modes. 

The rotation of the magnetized neutron star produces a difference in
the average streaming velocities of the plasma components. This
results in the circular polarization of the quasi--transverse modes for the
angles of propagation with respect to the local magnetic field less
than some critical angle. 
This may explain in a natural way the occurrence of the circular
polarization in the "core" emission.

If the cyclotron-Cherenkov instability 
occurs on the tail of the plasma distribution 
function then the particles of both signs of charge can resonate with the
wave. In a curved magnetic field, electrons and positrons will drift in
opposite directions  (Fig. \ref{driftDiff}). 
As the line of sight crosses the emission region,
the observer will
first  see the left circularly polarized wave emitted by the
electrons in the direction of their drift. 
When the line of sight becomes parallel to the local
magnetic field the wave will resonate with both electrons and positrons
in the plasma tail, so that the resulting circular polarization will be zero.
Finally, the observer will see the wave emitted by 
positrons in the direction of their drift. This can explain the switch of the
circular polarization observed in some pulsars.

The cone emission, which is due to the Cherenkov-drift instability
naturally has one linear polarization. An important difference from the 
standard
bunching theory is that the waves emitted at the Cherenkov-drift 
resonance are polarized perpendicular to the plane of the curved
magnetic field line. This may be used as a test to distinguish between the
two theories. To do so, one need to determine the absolute position
of the rotation axis of a pulsar - a possible but a difficult task.
One possible experiment would involve 
Harrison-Tademaru effect (\cite{HarrisonTademaru}),
 which 
predicts that the spins of neutron stars may be aligned with
their proper motion due to the quadrupole  magnetic radiation
if the magnetic moment is displaced from the center of the star.
Unfortunately, the current data does not support this
theory (in the Crab, the spin of the neutron star is within
$10^o$ of from the direction of the proper motion, while
for Vela it is approximately perpendicular).
Symmetry of plerions and direct observation of jets from pulsars 
(like the one observed in Vela pulsar (\cite{MarkwardtOgelman}))
combined with  polarimetry  of the pulsar
may be useful in determining the relative position
of the electric field of the wave and the magnetic axis.

\subsection{Radius-to-frequency mapping}

The cyclotron-Cherenkov mechanism predicts a simple dependence of the
emitted frequencies on the altitude (Eq. \ref{djjj1}).
The predicted power-law scaling of the emission altitudes is 
$R(\nu) \propto \nu ^{-1/6}$. This is strikingly close to the observed  
scaling $\propto \nu ^{-.15 \pm .1}$ (\cite{Lesch}, \cite{Kramer1994}), 
which is derived
from the simultaneous multifrequency observations of the time of arrival 
of pulses.
A simple "radius-to-frequency" mapping 
will be "blurred" by the scatter in energies of the 
 resonant particles, but the general trend that lower frequencies are 
emitted higher in the magnetosphere will remain.

The Cherenkov-drift instability, on the other hand, does not have 
a simple dependence of the emission altitudes on the frequency. 
The resonance conditions for the Cherenkov-drift instability are
virtually independent of frequency (Eq. \ref{aa11}), so that
the location of the emission region is determined by the various conditions
on the development of the instability (Appendix \ref{Conditiondrift}).

\subsection{Formation of  Spectra }

Development of the 
Cyclotron-Cherenkov and Cherenkov-drift instabilities 
results in an exponential growth of the 
electromagnetic waves.
 The original  growth may be limited by several factors.
A very likely possibility is that a spatial growth of the waves
is limited by the   changing parameters of the
plasma. This possibility is hard to quantify since we do not know
structure of the magnetic field in the outer regions of the
pulsar magnetosphere. 
The original  growth of the instabilities may also be limited 
by the nonlinear processes: quasi-linear diffusion, induced scattering
and wave decay.  Finally, 
 as the waves propagate in a
magnetosphere, they may be absorbed by the particles of the bulk plasma
(\cite{LyutikovPhD}). The emergent spectra are the combined
results of the these processes: emission, nonlinear saturation and absorption.

We have considered the two most likely  nonlinear saturation
effects for the cyclotron-Cherenkov instability: quasilinear diffusion  and 
induced Raman scattering (\cite{LyutikovRaman}, \cite{LyutikovQuasi}). 
In the case of quasilinear diffusion, the induced transitions
to upper Landau levels due the development 
of the  cyclotron-Cherenkov instability
is balanced by the  radiation reaction force due to the
cyclotron emission at the
normal Doppler resonance and the force arising in the inhomogeneous
magnetic field due to the conservation of the adiabatic invariant.
These forces result in a  quasilinear state and a 
saturation of the quasilinear diffusion.
We have found a  state, in which
the particles are constantly slowing down their parallel motion,
mainly due to the component along magnetic field  of the
radiation reaction force of emission at the
 anomalous
 Doppler resonance. At the same time they
 keep the pitch angle almost constant due to the balance
of the force arising in the inhomogeneous
magnetic field due to the conservation of the adiabatic invariant
 and
the component  perpendicular to the magnetic field  of the
radiation reaction force of emission at the
 anomalous
 Doppler resonance.
We calculated the distribution function and the wave intensities for
such quasilinear state (\cite{LyutikovQuasi}).

In the process of the quasilinear diffusion, the initial beam looses
a large fraction of its initial energy $ \approx 10 \%$, which
is enough to explain the typical luminosities of pulsars.
The theory predicts a spectral index $ F(\nu) \propto \nu ^{-2} $
($F(\nu)$ is the spectral flux density) which is very
close to the observed mean spectral index of $-1.6$
(\cite{lorimer}).
The predicted spectra also show a turn off at the low frequencies
$ \nu \leq 300 MHz$
and a flattering of the spectrum at large frequencies
$ \nu \geq 1 GHz$ (this may be related to the upturn in pulsar
spectra observed at mm-wavelength, 
(\cite{KramerXilouris})).

The other possible mechanism that may be  important for  wave
propagation and as an effective saturation mechanism for
instabilities of electromagnetic waves is 
induced Raman scattering of electromagnetic
waves  (\cite{LyutikovRaman}).
The frequencies, at which strong Raman scattering occur
in the outer parts of magnetosphere, fall into the observed
radio band.
The typical threshold
intensities for the strong Raman scattering are of the order
 of the observed intensities, implying that pulsar magnetosphere
may be optically thick to Raman scattering of electromagnetic waves.

Absorption processes can play an important role in the  formation of the
emergent spectra (\cite{LyutikovPhD}).
The waves may be strongly damped on the three possible resonances:
Cherenkov, Cherenkov-drift and cyclotron.
Alfv\'{e}n wave is always strongly damped on the Cherenkov resonances and possibly
on the Cherenkov-drift resonance and cannot leave magnetosphere.
Both ordinary and extraordinary waves may be damped on the
Cherenkov-drift resonance. In this case Cherenkov-drift resonance affects
only the ordinary and extraordinary waves with the electric field
perpendicular to the plane of the curved magnetic field line.
The high frequency parts of the
extraordinary and ordinary wave may also be damped on the  cyclotron
 resonances.
In fact, in a dipole geometry, electromagnetic waves propagating outward 
are always absorbed. The fact that we do see radiation implies that
 waves get "detached" 
from plasma escaping absorption.

\subsection{High Energy Emission }

It may be possible also to relate the pulsed high energy emission to the 
radio mechanisms. 
The development of the cyclotron-Cherenkov instability at the
anomalous Doppler effect leads to the 
finite pitch angles of the resonant particles. The particles will
undergo a cyclotron transition at the  normal Doppler effect 
decreasing the their pitch angle. The frequency of the wave emitted
in such a transition will fall in the soft  $X$-ray  range with
the frequencies 
\begin{equation}
\omega \approx \gamma_b \omega_B =  \, 10^{18} {\rm rad \, s^{-1} }
\hskip .2 truein (E \approx 1 {\rm keV})
\label{end}
\end{equation}
In such a model the high energy emission will coincide with the
core component of the radio emission, which is  what is 
observed in Crab pulsar and some other pulsars.
We also note, that this may be a feasible theory for the pulsed
 soft $X$-ray emission. The hard  $X$-ray and $\, \gamma$-emission
cannot be explained by this mechanism since the total energy flux
in the primary beam is not enough to account for the 
very high energy emission (see, for example, \cite{Usov1996}).

\section{Observational Predictions }
\label{Predict}

{\bf General Predictions of the Maser-type Instability}

The formalism of the maser-type emission assumes a random
phase approximation for the wave-particle interaction. Any given particle
can simultaneously  resonate with several waves 
whose phases are not correlated.
This is different from the reactive-type
 emission (like coherent emission by bunches) when the
intensities  of the  waves with different
frequencies are strongly correlated. Thus any observation
of the fine frequency structure in the pulsar
radio emission  may be considered as a strong argument 
against the reactive-type emission and in favor of the maser-type  emission.

{\bf Frequency Dependence of the Circular Polarization in the Core}

The cyclotron-Cherenkov instability can develop both on the 
primary beam particles and the particles from the tail of the 
plasma distribution. The cyclotron-Cherenkov instability on the
primary beam produces a pulse which has a maximum circular
polarization in its center. The cyclotron-Cherenkov instability on the
 tail particles  produces a pulse with the switching of the 
sense of the circular polarization in its center.
Since the resonance on the tail particles occurs on larger frequencies
( Eq. \ref{g1}) the effects of the switching 
sense of the circular polarization should be more prominent
on the higher frequencies.

{\bf Linear Polarization of the Cone Radiation}

Cherenkov-drift instability produces waves with the linear
polarization perpendicular to the plane of the curved field line.
This is in a sharp contrast to many other theories of the radio
emission that tend to generate waves with the electric field
in the plane of the curved field line.
If one can determine the absolute position of the rotational
axis,  magnetic moment and the electric vector of the  linear polarization,
then, assuming a dipole geometry,  it will be possible 
to determine  unambiguously the position of the  electric vector
of the emitted wave with respect to the plane of the magnetic
field line.

{\bf No  Cyclotron-Cherenkov Instability in  Millisecond Pulsars}

In  Appendix \ref{Conditionmillisecond} we show that the
 cyclotron-Cherenkov instability  does not develop in the 
magnetospheres of the millisecond pulsars. Since in our model
the  cyclotron-Cherenkov instability produces core-type
emission we predict that millisecond pulsars will not show 
core-type emission. At present, the existing observations do not
allow a separation into the core and cone type emission
in the millisecond pulsars.

\section{Conclusion}

In this work we presented a model for the pulsar emission generation
capable of explaining a broad range of
 the observations including the morphology,
polarization and spectrum formation in normal and millisecond pulsars.
We  provided no explanation for the temporal behavior, nulling, drifting subpulses although
these phenomena do not challenge the model. 
Our model makes several testable predictions, like high altitude
of the emission region and an unusual relation of the  polarization direction to the
magnetic axis for the cone
emission. In addition it has   flexibility to accommodate a variety of phenomena
  arising, for
example,
due to the temporal variations in the flow and the structure
of the magnetic field in a given pulsar,  or due to the
different structure of the  magnetosphere in different pulsars.

Further progress requires a better understanding
of the  structure of the pulsar magnetospheres both near the
stellar surface where the mutipole moments of the magnetic field can
substantially affect the  
physics of the pair formation front, and near the light cylinder, where the 
emission is taking place. It is also
  necessary  to consider the nonlinear stages of the emission mechanisms
described above.

\acknowledgments
We would like to thank George Melikidze for his comments.
ML would like to thank the Abastumani Astrophysics Observatory for
the hospitality during his stays in Tbilisi and GZM acknowledges the
support and  hospitality during his stay at Caltech.
This research was supported by NSF  grant AST-9529170.

\newpage

\appendix 

\section{Conditions on cyclotron-Cherenkov instability}
\label{Conditioncyclotron}

In this Appendix we consider the development of the
 cyclotron-Cherenkov instability in magnetosphere
of the typical pulsar. We show that the conditions on page
\pageref{conditions} (Section  \ref{Modelemission}) for the development
of the  cyclotron-Cherenkov instability instability are
 satisfied for  the typical pulsar.

The conditions for the development of the cyclotron-Cherenkov
instability may be easily derived for the small angles of propagation with
respect to the  magnetic field.
Representing the wave's dispersion as
\begin{equation}
\, \omega = kc (1 - \delta),\hskip .3 truein
\mbox{ where  } \delta \,=\, \left\{  \begin{array}{ll}
{\, \omega_p^2 T_p \over 4 \,  \gamma_p^3 \, \omega_B^2} & \mbox{ X mode} \\
{\, \omega_p^2 T_p \over 4 \,  \gamma_p^3 \, \omega_B^2} -
{\,  \gamma_p  \, \omega_p^2 T_p  \sin ^2 \theta   \over k^2 c^2 } 
& \mbox{ O mode} \\
{\, \omega_p^2 T_p \over 4 \,  \gamma_p^3 \, \omega_B^2}  +
{ k^2 c^2 \sin ^2 \theta  \over 4 \,  \gamma_p \, \omega_p^2 T_p} & \mbox{ Alfv\'{e}n mode}
\end{array} \right.
\label{aa10}
\end{equation}
and neglecting the drift term, the resonance condition (\ref{a}) may then
be written as 
\begin{equation}
{1 \over 2 \,  \gamma_{res}^2 } - \delta 
+ {  \theta ^2 \over 2} = \, -\,{\, \omega_B \over \omega \,  \gamma_{res} }
\label{g}
\end{equation}

Let us  discuss the condition for the development of the
cyclotron-Cherenkov instability. 
First we note that equation (\ref{g}) requires that
$ {1 \over 2 \,  \gamma_{res}^2 } \,<\,\delta$ and 
$  {  \theta ^2 \over 2} \,<\,\delta$. The first is the condition
that the particle is moving through plasma with the velocity faster than
the phase velocity of the wave. The  second  condition limits the 
emission to the small angles with respect to magnetic field. 
Assuming that $ {1 \over 2 \,  \gamma_{res}^2 } \,\ll \,\delta$ and
$  {  \theta ^2 \over 2} \,\ll \,\delta$ we find from (\ref{g})
\begin{equation}
\, \omega  = {\, \omega_B \over  \gamma_{res} \delta }
\label{g1}
\end{equation}
This resonant frequency increases with radius as $R^6$.

Using the upper limit on the frequencies for the relations  (\ref{aa10}) to 
hold, namely $\, \omega \ll  \,  \gamma_p \, \omega_B /T_p$. 
This sets the limit on $\delta$: $ \delta \gg T_p/(\,  \gamma_p \, 
 \gamma_{res})$. 
This condition is satisfied for the radii:
\begin{equation}
\left({R \over R_{NS} }\right) \,>\,
\left({ 2 \,  \gamma_p^2 \,  \omega_{B, NS} \over  \gamma_b \, \lambda \Omega}
\right)^{1/3} = 
\left\{ \begin{array}{ll}
 \,5 \times 10^2 & \mbox{ for the beam} \\
\, 2 \times 10^3 & \mbox{ for the tail}
\end{array} \right.
\label{g2}
\end{equation}
This is the radius where the resonance condition first becomes satisfied
for the particles of the primary beam and from the tail of the
distribution function. 

The cyclotron-Cherenkov instability growth rate  is 
 (e.g.,  \cite{Machabeli2}):
\begin{equation}
\Gamma =
 \sqrt{ { \pi \over 2}} {\, \omega _{p, res} ^2 \over \omega \, \Delta \gamma}
=   \sqrt{ {  \pi \over 2}} 
{ \, \lambda \, \lambda_{res} T_p \,  \gamma_{res} \over 
 \, \Delta \gamma \,  \gamma_p^3 } \, {\Omega^2 \over \omega_B}
\label{g3}
\end{equation}
where we have normalized the density of the resonant particles to the
Goldreich-Julian density 
 $\, \omega _{p, res} ^2 = \, \lambda_{res} \, \omega^2  _{GJ} $.

It follows from (\ref{g3}) that the growth rate increases with 
radius as $R^3$. Deeper in the magnetosphere the growth
rate is slow and the waves are not excited.
 At some point the growth rate becomes
comparable to the dynamic time $ \Gamma /\Omega = 1$. This occurs
at
\begin{equation}
\left({R \over R_{NS} }\right) = 
\left({  \, \Delta \gamma   \,  \gamma_p^3 
\over    \sqrt{ {  \pi \over 2 }} \, \lambda \, \lambda_{res} T_p
 \,  \gamma_{res}  } \, {\, \omega_B^{\ast} \over \Omega} \right)^{1/3}=
\left\{ \begin{array}{ll}
 \,2 \times 10^3 & \mbox{ for the beam} \\
\, 1 \times 10^3 & \mbox{ for the tail}
\end{array} \right.
\label{g4}
\end{equation}

Using the equipartition condition (\ref{v2}) we conclude that at a given
radius  
the growth on the tail particles is approximately the same (the  higher 
resonant
density is compensated by the higher resonant $\,  \gamma$-factor.

Starting this radius the waves will start to grow with the growth rate
increasing with radius. The highest frequency of the growing mode is 
determined by the condition (\ref{g1}) evaluated at the radius (\ref{g4}):
\begin{equation}
\, \omega_{max} =  \left\{ \begin{array}{ll}
 1\times 10^{8} {\rm   rad \, s^{-1} } 
 &\mbox{ for the resonance on the beam}\\
4 \times 10^{11} 
{\rm   rad \, s^{-1} }  &\mbox{  for the resonance on the tail }
\end{array} \right.
\label{g5}
\end{equation}
These estimates show, that for the chosen  plasma
parameters  the cyclotron-Cherenkov instability always 
develops on the tail of the distribution function and 
can also develop on the particles from the primary beam.
The higher density of the tail particles favors the 
development of the instability on the tail particles. 
The cyclotron-Cherenkov instability on the tail particles occurs
deeper in the magnetosphere on the higher frequencies than
the cyclotron-Cherenkov instability on the  primary beam, 
which can develop further out in the magnetosphere and 
produce emission at the lower frequencies.

The
growth rate of the cyclotron-Cherenkov instability
 should satisfy two other conditions:  the growth length must 
be  much less that the  length of the coherent
wave-particle interaction  and the condition of the kinetic
approximation. The coherent  wave-particle 
interaction  in limited by the curvature of the magnetic field.
A  particle can
resonate with the waves propagating in a limited range of angles
with respect to the magnetic field. As the wave propagates
 in a curved field the angle that the wave vector makes with the
 field changes and the wave may leave the range of resonant angles.
If  the range of the resonant angles is $\Delta \theta$ then
the condition of a short growth length is
\begin{equation}
{c \over \Gamma } < \Delta \theta R_B
\label{aa121}
\end{equation}

For cyclotron-Cherenkov instability we can estimate 
$\Delta \theta \approx \sqrt{ \delta}$.
Then we find a condition on the radius of curvature
\begin{equation}
R_B > {1\over \sqrt{ 2 \pi}}
 { c \, \Delta \gamma \over \lambda_{res} \Omega \,  \gamma_{res}
\delta^{3/2} } \approx 
\left\{ \begin{array}{ll}
10^{11} {\rm cm} & \mbox{ for the beam} \\
5 \times 10^{10} {\rm cm} & \mbox{ for the tail}
\end{array} \right.
\label{ab}
\end{equation}
at the distance $R = 2 \times 10^9 {\rm cm}$.
The region of the cyclotron-Cherenkov instability is limited to the
field lines near the  central field line. The transverse size of the 
emission region may be estimated as
 $x = R^2/R_B \approx  10^8 {\rm cm}$. This gives an opening
angle of the emission $ \theta^{em} = x/R \approx 2 ^{\circ}$.

There is another condition that the growth rate (\ref{g3}) should
satisfy, namely the condition of the kinetic approximation.
In deriving the growth rate we implicitly  assumed that
the wave-particle interaction is described by the random phase approximation,
which requires that
the spread of the resonant particles satisfies the
condition
\begin{equation}
\left| {\bf k} \delta {\bf v}_{res} \right| \gg \Gamma
\label{aa124}
\end{equation}

For the particle streaming  in the curved magnetic field
without any gyration
this condition takes the form
\begin{equation}
\left| k_{\parallel} c {\Delta \,  \gamma \over  \gamma^3} +
k_x u_d {\Delta \,  \gamma \over  \gamma}
 + {s \, \omega_B \Delta \,  \gamma \over  \gamma^2 } \right|
\gg \Gamma
\label{Gam}
\end{equation}

For the cyclotron-Cherenkov instability the last term on the left
hand side of (\ref{Gam}) is the dominant: 
\begin{equation}
{
\, \omega_B \, \Delta \gamma \over  \gamma_{res}^2 } \gg \Gamma
\label{Gam3}
\end{equation}

Using the growth rate (\ref{g3}) this condition may be rewritten
as 
\begin{equation}
\left({R \over R_{NS} }\right) \ll  
\left(\sqrt{ {2\over \pi}}
{ \, \Delta \gamma^2 \, \omega_B^{\ast ^3 }
\,  \gamma_p^3 \over \lambda_{res} \Omega^2 \,  \gamma_{res}
^3 \, \lambda T_p } \right)^{1/6}  = 3 \times 10^5
\label{g10}
\end{equation}
Equation (\ref{g10}) means that the kinetic approximation is
well  satisfied
inside the pulsar magnetosphere. 

Thus, the conditions for the development of the 
cyclotron-Cherenkov instability may be satisfied for the 
typical pulsar parameters. The waves are generated in the observed
frequency range near the central field line. The radius of emission
is $ \approx 10^9 {\rm cm}$, the transverse size of the emitting 
region is  $ \approx 10^8 {\rm cm}$ and the  "thickness" of the
emitting region is $\leq 10^9 {\rm cm}$. This may account for the 
core-type emission pattern.

\section{Grow rate for the Cherenkov-drift instability}
\label{growthrate}

In this Appendix we calculate the growth rate for the Cherenkov-drift
instability in the plane wave approximation.
Originally this growth rate has been obtained by
\cite{LyutikovMachabeliBlandford1}
using the single particle  emissivities. Here we give a derivation
using the simplified   dielectric tensor (\cite{LyutikovMachabeliBlandford1}).
The relevant components of the  antihermitian
part of the   dielectric tensor  are
(\cite{Kazbegi}).
\begin{eqnarray}
\epsilon^{\prime \prime}_{xx}&&= - i { 4 \pi^2 q^2 \over  \omega c }
\int {dp_{\phi} }   u_d^2 {\partial  f(p_{\phi} )  \over \partial p_{\phi} }
\delta\left( \omega- k_{\phi} v_{\phi} - k_x u_d \right),
\mbox{} \nonumber \\ \mbox{}
\epsilon^{\prime \prime}_{x \phi} &&   = - i { 4 \pi^2 q^2 \over  \omega c }
\int {dp_{\phi} }  u_d v_{\phi} {\partial  f(p_{\phi} )  \over \partial p_{\phi}
 }
\delta\left( \omega- k_{\phi} v_{\phi} - k_x u_d \right)
= \epsilon^{\prime \prime}_{\phi x} ,
\mbox{} \nonumber \\ \mbox{}
\epsilon^{\prime \prime}_{\phi \phi} &&   = - i { 4 \pi^2 q^2 \over  \omega  }
\int {dp_{\phi} }  v_{\phi}
{\partial  f(p_{\phi} )  \over \partial p_{\phi} }
\delta\left( \omega- k_{\phi} v_{\phi} - k_x u_d \right).
\label{qaq31}
\end{eqnarray}

The growth rate of the Cherenkov drift instability  can be obtained from
(\cite{Melrosebook1}):
\begin{equation}
\Gamma = \int d {\bf p}  w ( {\bf k} ,  {\bf p} ) \, \hbar  {\bf k}  \cdot
{\partial f ( {\bf p} ) \over \partial  {\bf p}  } .
\label{qaq3}
\end{equation}
The
growth rate for the lt-mode is
\begin{equation}
\Gamma^{lt} = { 4 \pi^2 q^2\over m } \int d p_{\phi}
\left( {k_{\phi} k_x u_d  \over c k k_{\perp}} - { v_{\phi} \over c} { k_{\perp}
 \over k} \right)^2
 { \partial f(p_{\phi}) \over  \partial p_{\phi} }
\delta\left( \omega- k_{\phi} v_{\phi} - k_x u_d \right),
\label{qaq7}
\end{equation}
and  growth rate for the t-mode is
\begin{equation}
\Gamma^{t} = { 4 \pi^2 q^2\over m } \int d p_{\phi}
\left( {k_r u_d  \over k_{\perp} c } \right)^2
 { \partial f(p_{\phi}) \over  \partial p_{\phi} }
\delta\left( \omega- k_{\phi} v_{\phi} - k_x u_d \right).
\label{qaq71}
\end{equation}

Where we chose the following  polarization vectors
\begin{eqnarray}
&&
{\bf e^{lt} } = {1\over k k_{\perp}   } \left\{ k_{\phi}  k_r, - k_{\perp}  ^2 ,
k_x k_{\phi} \right\},
\mbox{} \nonumber \\ \mbox{}
&&
{\bf e^t} ={1\over  k_{\perp}   } \left\{ - k_x ,0,  k_r,
\right\},
\label{qaq23}
\end{eqnarray}
where $k_{\perp}  = \sqrt{k_r^2 + k_x^2 } $
and 
\begin{equation}
{\bf k} =\left\{ k_r, k_{\phi} , k_x \right\} .
\label{qaq22}
\end{equation}
(see \cite{LyutikovMachabeliBlandford1}) for the discussion of the 
polarization properties of normal modes in anisotropic dielectric in 
cylindrical coordinates.

The maximum growth rate for the t-mode is reached when $k_x/k_{\phi} = u_d/c$
and the maximum growth rate for the lt-mode is reached when $k_r=0$.
We also note, that in the  excitation of both lt- and t-wave it is the
$x$ component of the electric field that is growing exponentially.

Estimating (\ref{qaq7}) and  (\ref{qaq71}) using $\delta$-function
( $ { \rm max } [ {k_r \over k_{\perp} } ]
\approx c \sqrt{2  \delta } /u_d$ and
max[$\left( {k_{\phi} k_x u_d  \over c k k_{\perp}} - { v_{\phi} \over c} 
{ k_{\perp}
\over k} \right)
$] $
\approx \sqrt{  2 \delta} $),
we wind the maximum growth rates of the t-
and lt-modes
in the limit $ \delta \gg 1/\gamma_{res}^2 $:
\begin{equation}
\Gamma ^{t} =  \Gamma ^{lt}  \approx  {2 \omega_{p, res}^2  \delta \over  \omega
 }
\left(
{\gamma^3 \over  1+ u_d ^2 \,  \gamma^2 /c^2  }
{\partial  f(\gamma)  \over \partial  \gamma } \right)_{res} ,
\label{qaq81}
\end{equation}
where $\omega_{p,res}$ is the plasma density of the resonant particles.

We estimate the growth rates (\ref{qaq81}) for the distribution
function of the resonant particles having a Gaussian form:
\begin{equation}
f(p_{\phi} )= { 1 \over \sqrt{2 \pi} p_t } 
 \exp\left( -{ (p_{\phi} -p_b)^2 \over 2 \Delta p^2}
 \right),
\label{qq1}
\end{equation}
where $p_b$  is the momentum of the bulk motion of the beam and
$\Delta p$ is the dispersion of the momentum. 

Assuming in (\ref{qaq81}) that $ u_d \gamma_b /c \gg 1$ we find
the growth rates
\begin{equation}
\Gamma ^{t} =  \Gamma ^{lt}  \approx \sqrt{ { 2\over \pi}}
  { \omega_{p, res}^2  \delta \gamma_b  \over  \omega  \Delta\gamma^2 }
{c^2 \over u_d^2} ,
 \label{qaq82}
\end{equation}
where $\gamma_b = p_b/( mc),\, \Delta \gamma= \Delta p/( m c)$.

\section{Conditions on Cherenkov-drift instability}
\label{Conditiondrift}

\begin{figure}
\psfig{file=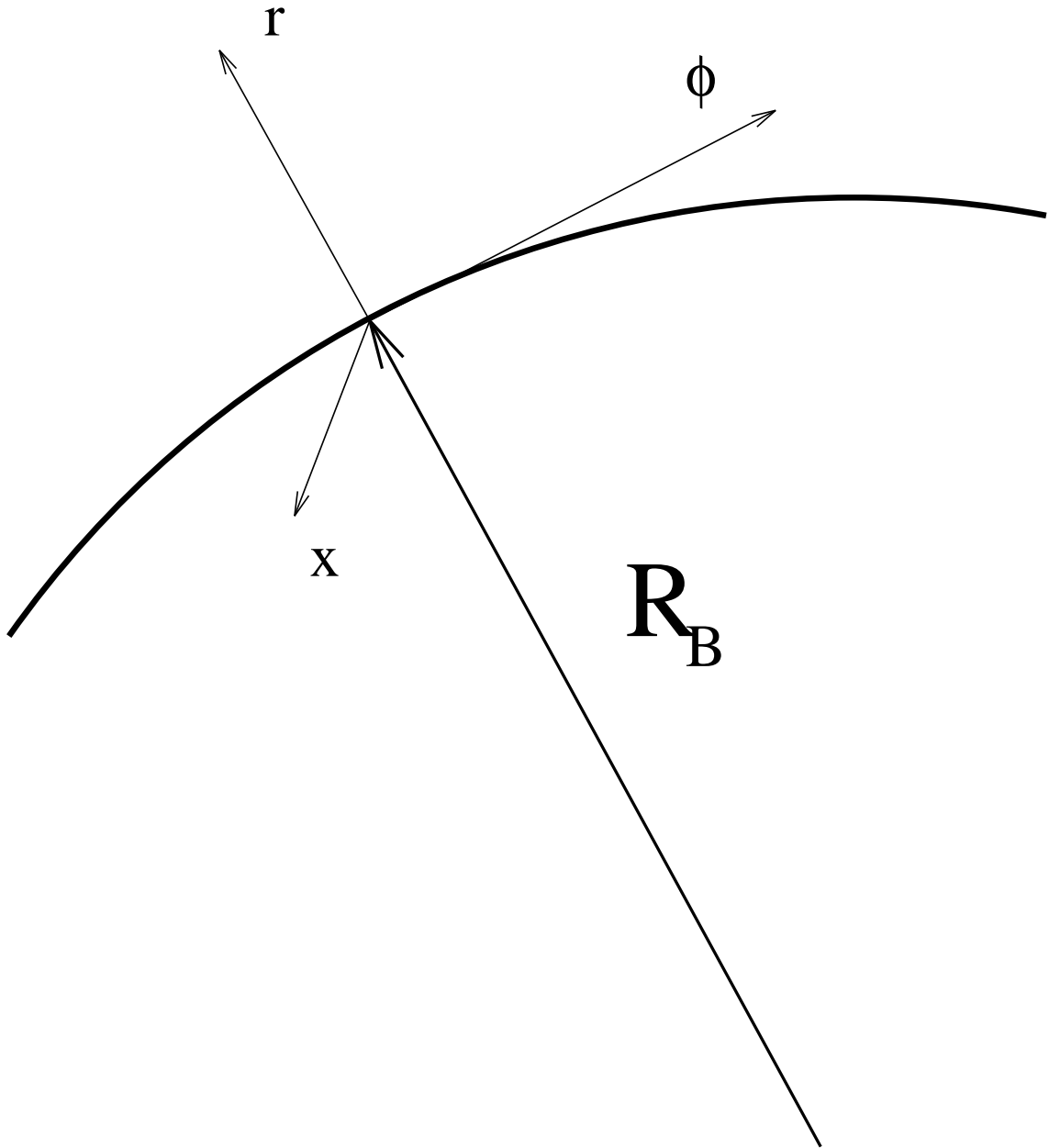,width=15.0cm}
\caption{
Coordinate system.
\label{Coordinate system}}
\end{figure}

Since Cherenkov-drift resonance requires a very high parallel momentum,
the resonant interaction will occur on the high phase velocity
waves. This implies that, like cyclotron-Cherenkov resonance,
   Cherenkov-drift resonance is always important for the
extraordinary mode, while for ordinary and Alfv\'{e}n  modes
it is important only for small angles of propagation.
We can then use the small approximation to dispersion relations for
the small angles of propagation (\ref{aa10}). Introducing
 cylindrical coordinates $x\,, r\,, \phi$  with
$r$ along the radius of curvature of the field line, $x$ perpendicular
to the plane of the curved field line and $\phi$ the azimuthal coordinate 
(Fig. \ref{Coordinate system}) 
the resonance condition (\ref{a1}) may then be written as
\begin{equation}
{1 \over 2 \,  \gamma_{res}^2 } - \delta +
{1 \over 2}  \psi ^2 +
{1 \over 2} \left( \theta - u_d/c \right)^2 = \,0
\label{aa11}
\end{equation}
where we used $v_{res}= c(1- {1 \over 2 \,  \gamma_{res}^2 } - {u_d^2\over 2 c}
) $, $  \theta = {k_x \over {k_{\phi}} }$
and $ \psi = {k_r \over {k_{\phi}} }$. 

We note that for the extraordinary mode $\delta$ is independent 
of frequency and  equation 
(\ref{aa11}) becomes  independent of frequency as well.
 This means that
for the
given angle of propagation Cherenkov-drift resonance
on the  extraordinary mode occurs
at all frequencies simultaneously. It is different from the 
cyclotron-Cherenkov resonance occurs at a particular frequency.
 For both ordinary and 
Alfv\'{e}n modes  $\delta$ does independent on  the frequency, so that
for a given angle of propagation the Cherenkov-drift resonance occurs at a fixed
frequency.

Let us now discuss the condition for the development of the 
Cherenkov-drift instability. First we drop the ${1 \over 2 \,  \gamma_{res}^2 }$
 term from equation (\ref{aa11}). This term is much smaller than $ \delta$ 
for the radii satisfying 
\begin{equation}
\left({ R \over R_{NS} } \right) \,>\,
\left({ \, \omega_B^{\ast} \, 
 \gamma_p^3 \over \Omega \, \lambda T_p \,  \gamma_b^2}  
\right)^{1/3}  \approx 1
\label{aa112}
\end{equation}
which is satisfied everywhere inside the pulsar magnetosphere.

\begin{figure}
\psfig{file=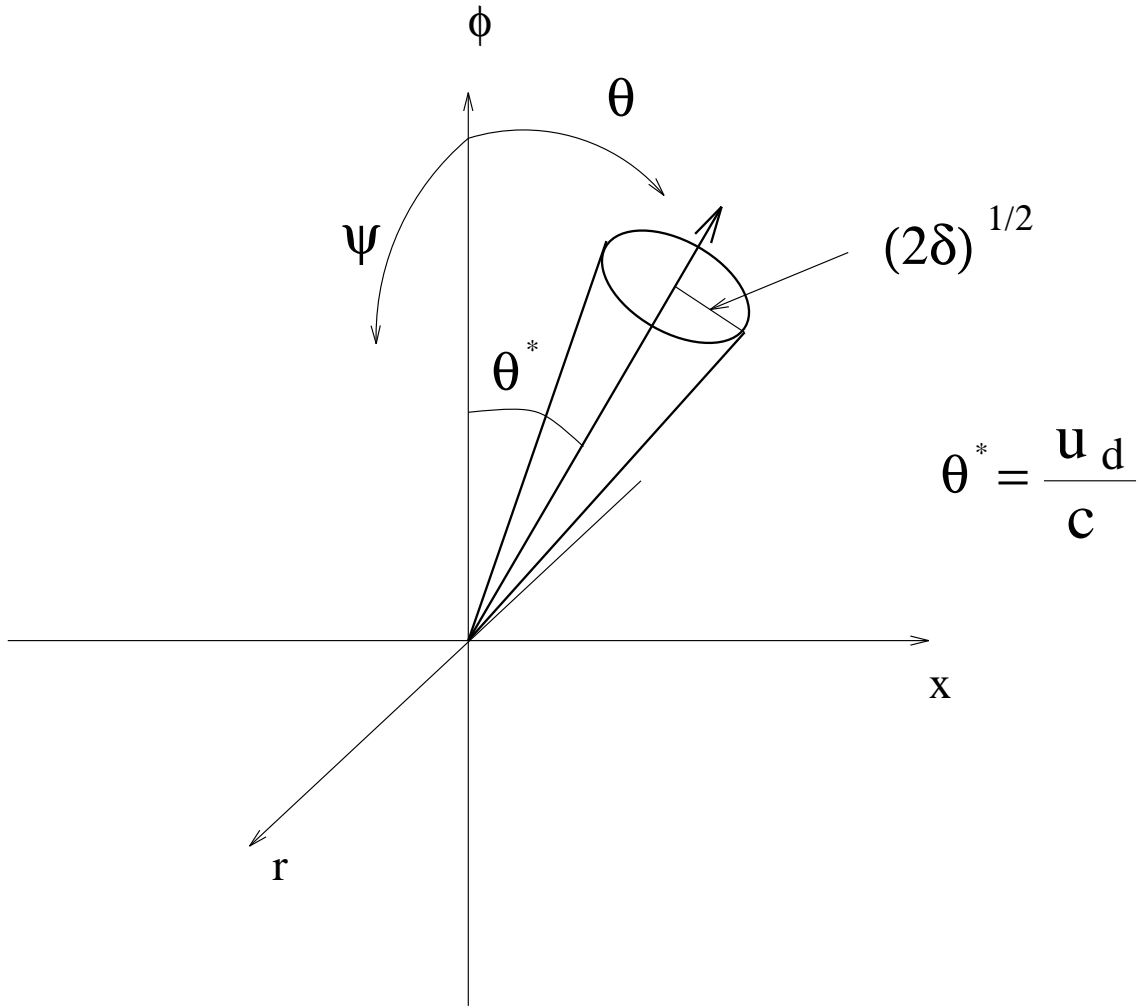,width=15.0cm}
\caption{ 
Emission geometry of the Cherenkov-drift resonance
\label{Cherenkovdrift1}}
\end{figure}

\begin{figure}
\center{
\psfig{file=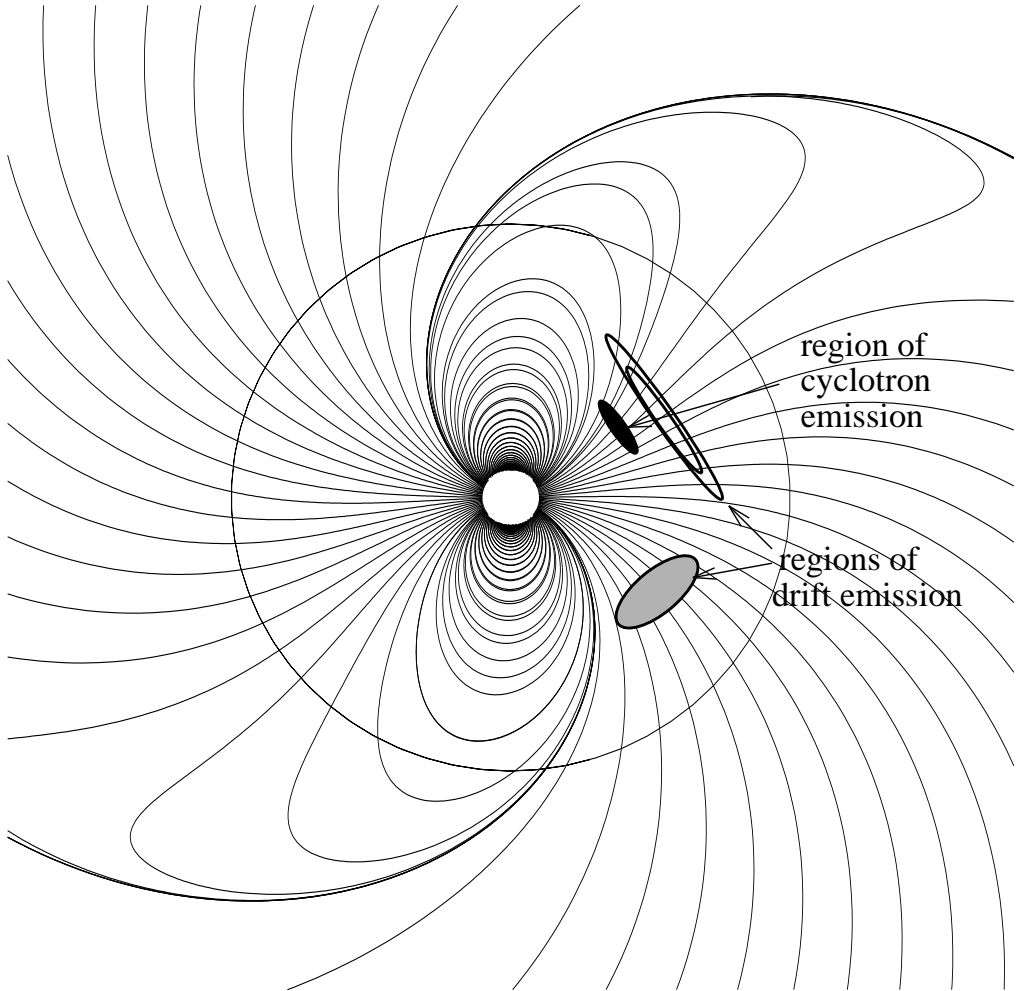,width=13cm}
\caption[Field of a rotating dipole in vacuum]
{Top view of the field lines in the equatorial
plane of a rotating point dipole in vacuum (\cite{YadigarogluP}).
Circle indicates light cylinder. The locations of cyclotron-Cherenkov
and drift instabilities are shown (similar regions will be on the other 
side of the pulsar). The
cyclotron-Cherenkov instability develops in the 
region of almost straight field lines. 
 The location of the Cherenkov-drift emission depends sensitively on the
curvature of magnetic field line. Two possible locations
of the Cherenkov-drift emission are shown: ringlike near the 
magnetic axis and in the region of swept field lines (shaded ellipse).
When 
 the effects of plasma loading are taken into account,
the field lines will become more curved. The current
flowing along the open field lines will also produce a torsion
in the field  lines, so that at the light cylinder the 
structure of the field will be changed considerably}}
\label{Deutsch}
\end{figure}

\begin{figure}
\psfig{file=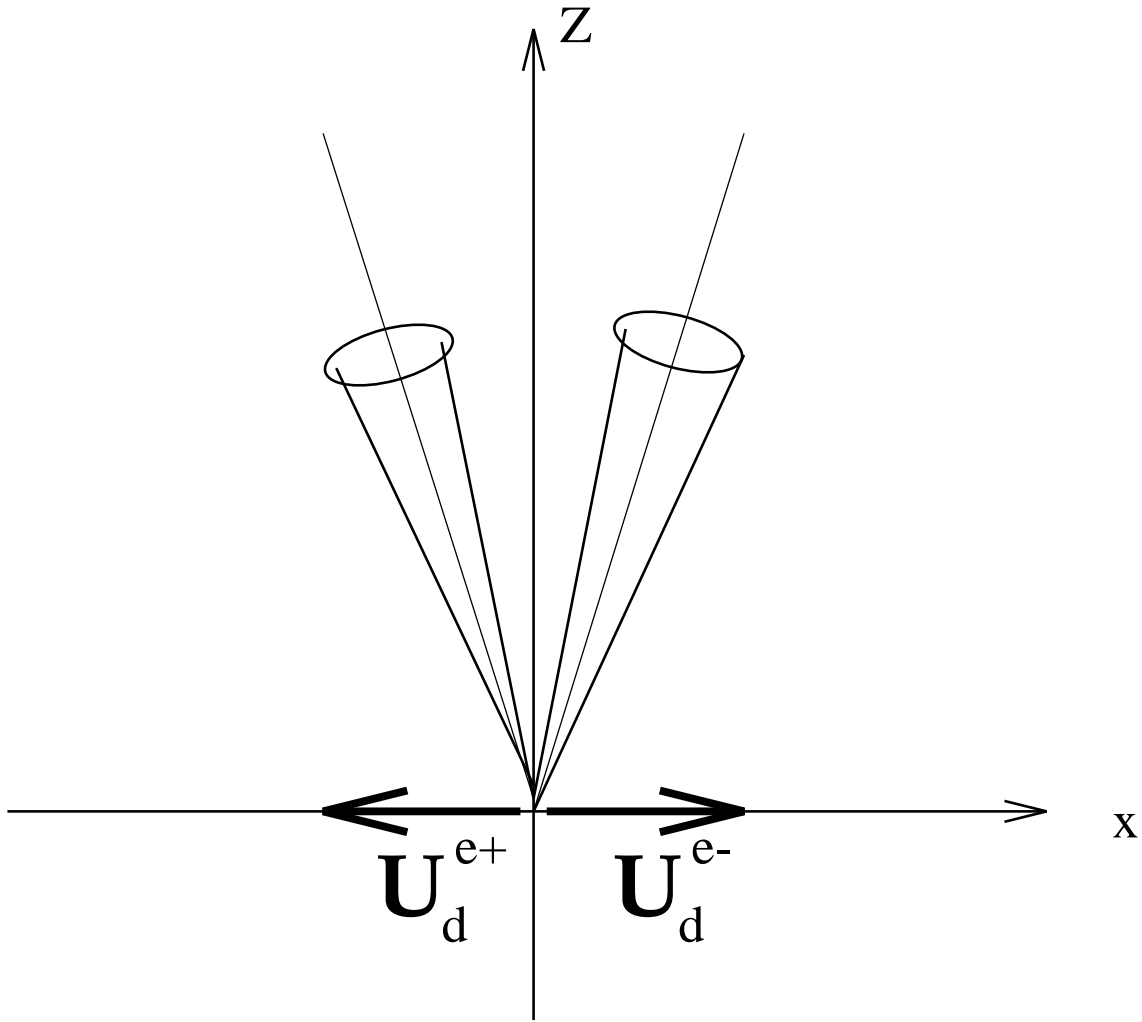,width=15.0cm}
\caption{
Switch of the sense of the circular polarization due to the
cyclotron-Cherenkov resonance on the tail of the plasma
distribution. The drift velocity of the electron and positron is 
in the opposite direction. As the line of sight crosses the emission region,
the observer first will see the waves emitted by the particles of one
sign of charge and then the other. The waves emitted by electron and 
positron have different circular polarization.
 \label{driftDiff}}
\end{figure}

We find then  that Eq. (\ref{aa11})
can be satisfied for 
\begin{equation}
 \psi \leq \sqrt{ 2\delta},
\hskip .2truein 
\left| \theta - u_d/c \right|  \leq \sqrt{ 2\delta}
\label{aa113}
\end{equation}
Fig. \ref{Cherenkovdrift1} describes the emission geometry of the 
Cherenkov-drift instability.

From (\ref{aa113}) we see that drift of the resonant particles
becomes  important if 
\begin{equation}
u_d/c > \sqrt{ 2\delta}
\label{aa114}
\end{equation}
Using the expression for $\delta$  (\ref{det3}), Eq.  (\ref{aa114}) gives
\begin{equation}
\left({ R \over R_{NS} } \right) \,>\, \left(
{ \, \lambda T_p   \Omega  \, \omega_B^{\ast} R_B^2 \over c^2 \,  \gamma_p^3 \, 
\, \gamma_b^2}
 \right)^{1/3}  
\label{aa115}
\end{equation}

At a given radius this condition may be considered as an upper limit
on the curvature of the field lines 
\begin{equation}
R_B \leq \left( {c^2 \,  \gamma_p^3 \,  \gamma_b^2 \over 
 \lambda T_p   \Omega  \, \omega_B } \right)^{1/2} \approx 
2 \times  10^9 {\rm cm} \, {\rm for}\, R = 1000 \, R_{NS}
\label{aa1151}
\end{equation}

Alternatively, condition (\ref{aa115}) may be regarded as a lower limit
on the radius from the star.
In the dipole geometry the radius of curvature for the
open field lines may be estimated as $R_B \geq  \sqrt{{  R \, c  \over \Omega}}$
The 
we find from  Eq.  (\ref{aa115})
\begin{equation}
\left({ R \over R_{NS} } \right) \,>\, \left(
{ \, \lambda T_p  \, \omega_B^{\ast} R_{NS}  \over c  \,  \gamma_p^3 \,  \gamma_b^2}
 \right)^{1/2}
= 23
\label{aa117}
\end{equation}
 In what follows we assume that the Cherenkov-drift resonance
occurs in the outer parts of the pulsar magnetosphere where the typical
value of the drift velocity is $ u_d \approx 0.01 c$.

The growth rate for the Cherenkov-drift resonance instability on the
primary beam is
calculated in appendix \ref{growthrate}, Eq. \ref{qaq82}.
Expressing the growth rate \ref{qaq82} in terms of the parameters
of the magnetospheric plasma we find
\begin{equation}
\Gamma= \sqrt{ { 2\over \pi}}
{ \lambda T_p  c^2 \over \gamma_p^3 \Delta \gamma^2 u_d^2}
{\Omega ^2 \over \omega}
\label{qaq8201}
\end{equation}

The
growth rate (\ref{qaq8201}) should satisfy several conditions.
The first is the criteria for the fast growth: $\Gamma / \Omega > 1$.
This is the requirement that the growth rate is fast enough, so that the
instability can have time to develop before the plasma is carried out of the 
magnetosphere. 
From (\ref{qaq8201}) we find
that  the condition $\Gamma / \Omega > 1$ 
is satisfied for the chosen parameters and 
 the typical frequency of emission
 $\, \omega= 5 \times 10^9$   rad \, s$^{-1}$.

The next condition that a growth rate should satisfy is that the
growth length be much less that the  length of the coherent 
wave-particle interaction (\ref{aa121}). 
Estimating the  range of resonant angles $\Delta \theta \approx \sqrt{\delta}$
 and using the growth rate (\ref{qaq8201}), this condition gives
\begin{equation}
R_B \geq { c \, \omega \over  \omega _B \Omega } \,
{\, \Delta \gamma^2  \over  \gamma_b \delta^{3/2} } 
\approx 10^{10} {\rm cm} \mbox{ at $R \approx 10^9 {\rm cm}$}
\label{aa122}
\end{equation}

Equation (\ref{aa122}) is the lower limit on the radius of curvature of the
field lines. It will restrict the emission region to the field lines
closer to the central field line, where the  radius  curvature  is large.

There is a third condition that the growth rate (\ref{qaq8201}) should 
satisfy -  the condition of the kinetic approximation  (\ref{Gam}).
For the Cherenkov-drift resonance condition (\ref{Gam}) gives
\begin{equation}
\left| k_x u_d {\Delta \,  \gamma \over  \gamma}\right| \gg \Gamma
\label{Gam1}
\end{equation}
Estimating $k_x c \approx \,  \omega u_d/c $ we find
that this condition can be easily satisfied in the pulsar magnetosphere.

In this Appendix we showed that the Cherenkov-drift instability can
develop in the pulsar magnetosphere. The region of the development of the 
instability has a radius of curvature limited both from below
(by the  condition of a large radius of curvature \ref{aa1151})
and from above (by the condition
of a short growth length  \ref{aa122}).

\section{ Instabilities in millisecond pulsars}
\label{Conditionmillisecond}

Both cyclotron-Cherenkov and Cherenkov-drift instabilities may develop
in  millisecond pulsar.  It is harder to satisfy the conditions
for the development of the cyclotron-Cherenkov instability
in the  millisecond pulsars than in the normal pulsars. Since there is no
clear distinction between core and cone-type emission for the 
 millisecond pulsars it is  possible that only Cherenkov-drift instability
develops in their magnetospheres. Alternatively, different
pulsars may have a substantially different plasma parameters, so that
both instabilities may develop in  some of them.

 Here we will discuss briefly the conditions 
for the development of thes instabilities in a "standard"  millisecond pulsar
with the period $P= 5 \times 10^{-3} $ s and the  surface magnetic field 
$B=10^8$ G. As a first order approximation we will assume that the 
other plasma parameters, i.e. the  initial primary beam Lorentz factor
$\, \gamma_b ^{(0)}= 6 \times 10^7$, 
the  primary beam Lorentz factor at the light cylinder
$\, \gamma_b ^{(0)}= 6 \times 10^6$, 
its  scatter $\Delta \gamma=10^2$,
 average streaming
energy of the secondary plasma $\, \gamma_p=10$, its scatter in energy $T_p=10$ 
and the
multiplicity factor $\lambda= 3 \times 10^5$ are the same. 
The light cylinder is now at a radius $R_{lc}= 2.4 \times 10^7 {\rm cm} =
24 \, R_{NS}$.
These are very approximate assumptions. The plasma 
parameters in the millisecond pulsars are likely to be different from
the normal pulsars. At this moment we do not understand the 
physics of the pair production well enough to make a quantitative 
distinction between normal and millisecond pulsars. 

The growth rate of the cyclotron-Cherenkov instability
becomes equal to the rotational frequency of the pulsar at
(see (\ref{g4}))
 \begin{equation}
\left({R \over R_{NS} }\right) = 20
\label{h}
\end{equation}
which is  inside the light cylinder.

The approximation of the kinetic growth rate
 cyclotron-Cherenkov instability (\ref{g10}) requires that
\begin{equation}
\left({R \over R_{NS} }\right) < 3 \times 10^3 
\label{h1}
\end{equation}
which is satisfied.
The condition of a short growth length of the cyclotron-Cherenkov instability 
 (\ref{aa121}) requires that
\begin{equation}
R_B \geq 10^{9} {\rm cm}
\label{h2}
\end{equation}
at the emission site of $R \approx 20 \, R_{NS}$.
This condition is hard to satisfy: 
 cyclotron-Cherenkov instability does not develop
in the millisecond pulsars.

For the Cherenkov-drift instability, the condition of a large
drift (\ref{aa114}) is  now  satisfied for the all open  magnetic field lines
as well as the 
 conditions of a fast growth  (condition (i) on page \pageref{gr})
ans the condition of the kinetic approximation  (condition (ii) on page \pageref{gr}).
Cherenkov-drift
instability does 
develops in the magnetospheres of the  millisecond
pulsars.

\section{ Effects of curvature radiation reaction on the beam}
\label{cooling}

Consider a beam of electrons propagating in s dipole curved magnetic
field. The radius of curvature on a field line near the 
the magnetic moment is 
\begin{equation}
R_B = { 4\over 3} {\sqrt{ R_{NS} R } \over  \alpha ^{\ast}}
\label{f45}
\end{equation}
Here $ \alpha ^{\ast} \ll \sqrt{ R_{NS} /R} $ 
is the angle at which a given field lines intersects the neutron star
surface.

An evolution of the distribution function under
the influence of the radiation reaction form is described by the 
equation:
\begin{equation}
{\partial f (z,p_z,t) \over \partial t} +
 {\partial  \over \partial p_z} 
\left({\partial  p_z \over \partial t} f (z,p_z,t) \right)
=0
\label{f451}
\end{equation}

Equation  (\ref{f451}) can be solved by integrating along characteristics
\begin{equation}
{\partial p_z \over \partial t} = 
-{ 2 e^2 \, \gamma^4 v_z^3 \over 3 c^3 R_B^2}
\label{f454}
\end{equation}

With the radius of curvature given by  (\ref{f45})
Equation (\ref{f454}) has a solution

\begin{equation}
{1\over 2} \left(
{ \sqrt{ p_z^2 +(m c)^2} \over  p_z^2 } +
\ln \left({  p_z \over 1+  \sqrt{ p_z^2 +(m c)^2} }\right)  \right)
\left. \phantom{ {{{ {a\over b} \over {a\over b} } } \over {{ {a\over b} \over
{a\over b} } } } } \right|_{p_0}^ {p_z(t) }=
-{ 3 \over 2} { r_e  \alpha ^{\ast 2}  \over R_{NS} } \ln (t/t_o) 
\label{f455}
\end{equation}
where $r_e = { e^2 \over m c^2} $ is the classical radius of an electron,
 $t_0 \approx R_{NS}/c$ and $p_0$ is the initial momentum.
This equation can be  solved for $ p_z \gg m c $
\begin{equation}
\, \gamma_0 ( \, \gamma,t) =
 \, \gamma \, \left(1- \, \gamma^3 { 9 r_e \alpha ^{\ast 2} \over 2
 R_{NS}
} \, \ln(t/t_0) \right)^{-1/3}
\label{f47}
\end{equation}

\begin{figure}
\psfig{file=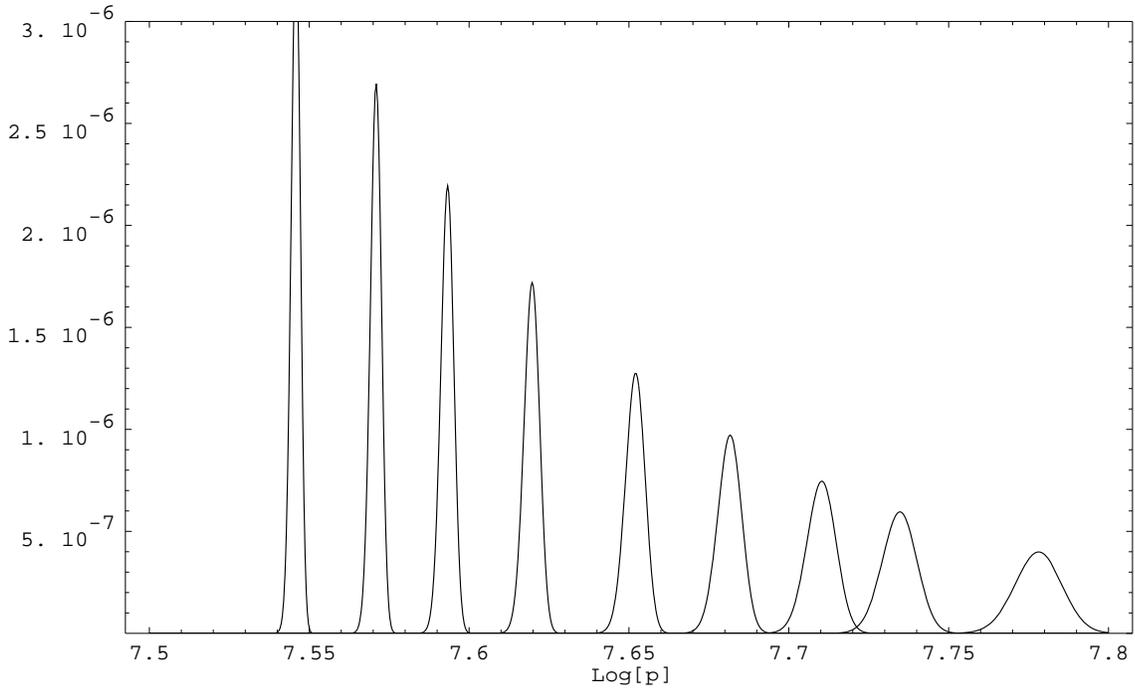,width=15.0cm}
\caption{
Evolution of the primary beam distribution function due the curvature
radiation reaction. For the illustrative purposes we chose the 
initial  beam $\, \gamma$-factor equal to  $6 \times 10^7$,
and the initial scatter in Lorentz factors $\Delta \gamma =10^6$. 
\label{distribution1}}
\end{figure}

Solution of the continuity equation (\ref{f451}) is then
\begin{equation}
 f (z,\, \gamma,t) = f_0 \left(  
\, \gamma_0(\, \gamma,t), \phantom{  { {a\over b} \over
{a\over b} } } t=t_0 \right)
\, \left({ \, \gamma \over \gamma_0(\, \gamma,t) } \right)^ {-4}
\label{f471}
\end{equation}
where $ f_0 (\, \gamma_0, t=t_0) $ is the initial distribution
function at $R= R_{NS}$.
Evolution of the distribution function is shown in Fig. \ref{distribution1}.

If, originally,  the beam had a scatter in energy $\Delta \gamma_0$ then 
at a later moment it will have a scatter in energy given by
\begin{equation}
\Delta \gamma = \Delta \gamma_0 \,
\left(1+ \, \gamma_0^3 { 9 r_e \alpha ^{\ast 2} \over 2 R_{NS}
} \, \ln(t/t_0) \right)^{- 4/3} 
\label{f48}
\end{equation}

Which implies that 
the scatter in energies of the beam may decrease much faster than its
average energy.

To estimate the decrease in the average energy and in the energy  scatter
onsider an evolution of the beam on the last open field line. Then
$\alpha ^{\ast 2} =  R_{NS} \Omega / c$ and at the light cylinder 
($t =  2 \pi / \Omega$) we have
\begin{eqnarray}
&&
 { \, \gamma \over \gamma_0} =
\left(
1+ \, \gamma_0^3  { 9\over 2} { r_e \Omega \over c} 
\ln ({ 2 \pi c\over R_{NS} \Omega })  \right)  ^{-1/3} 
\mbox{} \nonumber \\ \mbox{}
&& {\Delta \gamma   \over \Delta \gamma_0} =
\left(
 1+ \, \gamma_0^3  { 9\over 2} { r_e \Omega \over c}
\ln ({ 2 \pi c\over R_{NS} \Omega }) \right)  ^{-4/3} 
\label{f49}
\end{eqnarray}
Which may be expressed as
\begin{equation}
{ \Delta \gamma \over \gamma} =
 { \Delta \gamma_0 \over   \gamma_0} \,
\left( {  \gamma  \over \gamma_0} \right)^3 
\label{f491}
\end{equation}
If the original beam was mildly relativistic ( $ \Delta \gamma_0 \approx
0.1 \, \gamma_0$ then in order to reach a scatter $ \Delta \gamma/\, \gamma
\approx 10^{-4}$ the beam have to lose about 90\% of its original
energy: $\, \gamma / \, \gamma_0 \approx 0.1$.
Estimating (\ref{f49}) with   we find that  the scatter in the 
energy of the beam on the last open field line at the light cylinder 
for the normal pulsars may be as small as
$ \Delta \gamma_{lc} =  100 $. 
This vindicates our assumption of the chosen
beam energy spread.

In the normal pulsars the cooling of the high energy beam is important for the
$\, \gamma_0 \geq 2 \times 10^7$. In  millisecond pulsars it becomes
 important for 
$\, \gamma_0 \geq  2 \times 10^6$. 
The cooling of the resonant particles
increases  the growth rates of both cyclotron-Cherenkov and 
Cherenkov-drift instabilities. On the other hand, the kinetic approximation
condition for the Cherenkov-drift instability may not be satisfied for 
a very small scatter in energy of the resonant particles. 
Unlike the cooling, the change in the average energy of the beam is normally
not very important and is neglected.

\section{Instability in an ion beam}
\label{Instabilityion}

If the relative orientation of rotational and magnetic axis is such that the electric
field pulls up positive charges, then the primary beam will consist of ions. 
The critical Lorentz factor of the charge particles needed for the pair 
production is independent of the mass of the particle. Ions will be accelerated
to the  same Lorents factors as electrons $\gamma_b \approx 10^{6-7} $.
It is natural to expect that in approximately $50\%$ of  neutron stars the
primary beam will consist of ions. 
In what follows,
the ratio of  ion and electron masses will be denoted as
$\rho= m_i/m_e$.

First we consider a cyclotron-Cherenkov instability on the ion
beam.
If the primary beam consists of ions, then the  cyclotron-Cherenkov resonance
at the given frequency occurs closer by a factor $\rho ^{1/6} \approx  3$ 
(see \ref{g2}). This will
increase the density of the resonant particles by $\approx 30$, but the growth
rate for the  cyclotron-Cherenkov  instability (\ref{g3}) 
is  proportional to the squared 
plasma
frequency of the resonant particles, which is inversely proportional to the mass
of the particles.
 Taking into account that the condition of the kinetic instability
 is independent of the 
mass of the particle,
we conclude that the  cyclotron-Cherenkov  instability  on the ion
beam does not develop due to the very  small growth rate.

The growth rate of the Cherenkov-drift instability is proportional to 
$\rho ^{-3}$ (Eq. \ref{qaq82}), so the  growth rate on the ion
 beam is many orders of magnitude smaller than the  growth rate on the
electron beam.
Thus, we conclude that both  cyclotron-Cherenkov and Cherenkov-drift 
instabilities do not develop on the ion beam.
The pulsars with such a relative orientation of magnetic and 
rotational axis that the induced electric field accelerates 
ions can produce radio emission only at the 
cyclotron-Cherenkov instability on the tail particles.

\newpage

\newpage


\end{document}